\documentclass[12pt]{article}
\pdfoutput=1
\usepackage[utf8]{inputenc}
\usepackage[T1]{fontenc}
\usepackage{float}
\usepackage{authblk}
\usepackage[normalem]{ulem}
\usepackage{xcolor}
\usepackage{amsmath}
\usepackage{amsthm}
\usepackage{amssymb}
\usepackage{amsfonts}
\usepackage{adjustbox}
\usepackage{bbm}
\usepackage{graphicx}
\usepackage{booktabs}
\usepackage{mathtools}
\usepackage{multirow,array}
\usepackage{fullpage}
\usepackage{comment}
\usepackage[mathlines]{lineno}
\usepackage{multirow}
\usepackage{subcaption}
\usepackage[style=numeric-comp,sorting=none,maxnames=99]{biblatex}

\DeclareMathOperator*{\argmax}{arg\,max}

\usepackage[colorlinks = true,
            linkcolor = teal,
            urlcolor  = teal,
            citecolor = teal]{hyperref}

\title{An analytical model of active inference in the\\Iterated Prisoner's Dilemma}

\author[1,2]{Daphne Demekas\thanks{\href{mailto:daphnedemekas@arizona.edu}{daphnedemekas@arizona.edu}}}
\author[1,3,4,5]{Conor Heins\thanks{\href{mailto:cheins@ab.mpg.de}{cheins@ab.mpg.de}}}
\author[1,5,6]{Brennan Klein\thanks{\href{mailto:b.klein@northeastern.edu}{b.klein@northeastern.edu}}}

\affil[1]{Network Science Institute, Northeastern University, Boston, Massachusetts, USA}
\affil[2]{Wheeler Lab, University of Arizona, Tucson, Arizona, USA}
\affil[3]{Department of Collective Behaviour, Max Planck Institute of \protect\\Animal Behavior, 78464 Konstanz, Germany}
\affil[4]{Department of Biology and the Centre for the Advanced Study of Collective Behaviour,\protect\\University of Konstanz, 78464 Konstanz, Germany}
\affil[5]{VERSES AI Research Lab, Los Angeles, CA 90016, USA}
\affil[6]{The Institute for Experiential AI, Northeastern University, Boston, Massachusetts, USA}

\begin{document}
\maketitle
\pagenumbering{arabic}

\begin{abstract}
This paper addresses a mathematically tractable model of the Prisoner's Dilemma using the framework of active inference. In this work, we design pairs of Bayesian agents that are tracking the joint game state of their and their opponent's choices in an Iterated Prisoner's Dilemma game. The specification of the agents' belief architecture in the form of a partially-observed Markov decision process allows careful and rigourous investigation into the dynamics of two-player gameplay, including the derivation of optimal conditions for phase transitions that are required to achieve certain game-theoretic steady states. We show that the critical time points governing the phase transition are linearly related to each other as a function of learning rate and the reward function. We then investigate the patterns that emerge when varying the agents' learning rates, as well as the relationship between the stochastic and deterministic solutions to the two-agent system.
\end{abstract}

\section{Introduction}

Studies of behavioural science, be it in biology, psychology, or machine learning, often rely on the concept of rational thinking and decision making \cite{axelrod1981evolution, nowak2006five, nowak1993strategy, simon1990bounded}. Game theory has had wide success in precisely formulating contexts in which players or agents are challenged to converge to an optimal yet counter-intuitive strategy that maximises reward. In particular, game theory models communication among agents that can result in bounded-complex emergent behaviour \cite{gametheory, Vukov2006cooperation}. The Iterated Prisoner's Dilemma (IPD) is a quintessential game, in which the `dilemma' is that the highest reward is attributed to the action of defection, but the optimal behaviour in the long run is to cooperate, because of the `Shadow of the Future' phenomenon \cite{shadow}\footnote{This is when agents in repeated play---without awareness of when the play will end---will be more cooperative because they are made to learn about the possibility of being punished and plan accordingly \cite{pd}.}. When played iteratively, agents learn each other's predictable behaviour and can form an optimal strategy, away from the Nash equilibrium of the one-shot game. To do so, agents need to be aware of what their opponent is likely to do, which is why the IPD is widely used to study the evolution of cooperation for selfish agents \cite{pavlov}.

This work addresses a computational model of the (memory-one) Iterated Prisoner's Dilemma under the framework of active inference (AIF) \cite{Ramstead2022, Heins2022spin, parr2022a}. AIF is an agent-based modelling framework derived from theoretical neuroscience, where cognitive processes like action, perception, and learning are seen as solutions to an inference problem. As an explicitly model-based, Bayesian framework for simulating behaviour, AIF provides cognitively `transparent' agents, whose posterior beliefs about the world and associated uncertainties are accessible and interpretable. This enables careful investigation into the Bayesian basis of behaviour in these simple models, in turn allowing us to identify the conditions under which optimal behaviour is possible.

When two identical and deterministic AIF agents play against one another, we show that the equation governing across-trial learning dynamics is mathematically tractable given one approximation. This enables us to derive functions that model the specific conditions under which convergence to an optimal strategy---namely the Pavlov Strategy \cite{pavlov}---for the IPD can occur, given a multi-agent AIF model. The Pavlov strategy is win-stay-lose-change, where agents will cooperate if the agent's and opponent's moves are the same in the previous round and defect otherwise. We explore how these dynamics vary across different configurations of the agents' learning rates, as well as how stochasticity in the agent network determines the probabilities of agents reaching the optimal outcome.

\subsection{Iterated Prisoner's Dilemma}

In the Prisoner's Dilemma, at each round, both players can either defect or cooperate, leading to 4 possible outcomes \cite{pd} (see Table \ref{table:rewards} with different reward levels). The outcome with the highest reward is if the player defects and its opponent cooperates (DC), which is also the outcome with the lowest reward for the opponent (CD). The second-best outcome is if both cooperate (CC), and the third-best outcome for both players is if they defect (DD). In this model, the four reward levels are respectively [3,1,4,2]. This work specifically models the memory-one IPD, where each player only considers the previous move of their opponent when making their decision for the current round.

\begin{table}[t]
    \centering
    {\renewcommand{\arraystretch}{1.5}
    \begin{tabular}{cc|c|c|}
        & \multicolumn{1}{c}{} & \multicolumn{2}{c}{\textit{\underline{Player 2}}}\\
        & \multicolumn{1}{c}{} & \multicolumn{1}{c}{Cooperate (\textbf{C})} & \multicolumn{1}{c}{Defect (\textbf{D})} \\\cline{3-4}
        \multirow{2}*{\textit{\underline{Player 1}}} & \textbf{C} & $(3,3)$ & $(1,4)$ \\\cline{3-4}
        & \textbf{D} & $(4,1)$ & $(2,2)$ \\\cline{3-4}
    \end{tabular}}
    \caption{\textbf{Example payout matrix in a Prisoner's Dilemma game.}}
    \label{table:rewards}
\end{table}

There are several notable strategies in the IPD, which have been categorised in different ways \cite{strategies}. First, a dominant strategy produces the best possible payoff for an agent, regardless of the strategies used by opponents. The most commonly cited dominant outcome is when both players defect (choose to betray) in every round. From an individual player's perspective, defecting in every round provides a higher immediate payoff compared to cooperation, especially when the other player cooperates. However, defecting in every round is not socially optimal as it leads to a lower overall payoff compared to mutual cooperation. The challenge is to find strategies that can foster cooperation and lead to better outcomes for both players in the long run, rather than succumbing to the dominant outcome of mutual defection \cite{nash}.

In order to reach the social optimum of cooperation, new heuristics or bounds on the agents need to emerge in order for them to look beyond the reward function when deciding their actions. This makes the IPD a good arena to study \textit{bounded} rationality, in which agents do not have access to the full generative process (encompassing both themselves and their opponent), and therefore must make decisions given a bound on their awareness or knowledge, of, for instance, the other player, or any external environmental factors.\cite{simon1990bounded}. Agents playing the IPD has been studied in the context of reinforcement learning already \cite{rl, lin2022online}, and the idea of bounded rationality serves as a motivation for using active inference agents to model the IPD, as the AIF is a transparent and interpretable framework in which agents infer actions and quantify uncertainty under the constraints of their generative model.

There are several ways to train the agents to converge to the social optimum, which we will refer to as the cooperative steady state. When agents sample their actions deterministically, our model shows that active inference agents parameterised with a constrained set of learning rates can converge to the cooperative steady state by learning the Pavlov Strategy \cite{pavlov}, and it also demonstrates learning rate configurations that get trapped in the Nash equilibrium, in which agents converge to Unconditional Defection \cite{Press2012, Sandholm1996}.

\subsection{Active inference}

Active inference (AIF) agents are able to plan and learn about their state space and transition probabilities through observed experience. They infer which actions to take by minimising the expected free energy anticipated to accrue from their actions \cite{parr2022a}. This often allows these agents to solve complex tasks often seen in reinforcement learning or neuroscience, such as the Multi-Armed Bandit \cite{mab} and other Monte-Carlo based tasks \cite{Fountas2020deep}.

Advances in the ability to quickly build and scale models of AIF agents, particularly in Python using the \texttt{pymdp} library \cite{Pymdp2022}, have allowed for a much more scalable and accessible means to model these agents in different and flexible environments, as well as to connect them in networks and allow them to observe each other's actions. This has allowed researchers to ask more interesting questions about how relevant AIF is in terms of modelling rational decision making, such as those observed in game theory. In this paper, we show that not only can AIF agents effectively learn optimal strategies to the IPD, but the framework of active inference enables us to derive the exact conditions for when this will occur and have a layered understanding of the agents' `mental process' throughout the game.

The agents in this model actively entertain beliefs about the dynamics of the game and iteratively update their beliefs about the game dynamics (i.e., a `transition model') as they play multiple rounds against their opponent. In the context of the discrete-time and -space models used in the present work, this amounts to updating the elements of transition probability matrices that represent each agent's beliefs about game states from one trial to the next. After every trial of iterated play, the agents update these state transition probability distributions based on their actions and the outcomes that they observed. In doing so, the agents have the capacity to learn strategies, manifested as patterns of learned probabilities of transition from each state to each other state.

Our hypothesis is that throughout iterative play, the bounded-rational agents will learn to infer actions based on learned patterns of their opponent's behaviour (i.e., the ability to predict revenge from defection), and this will result in a strategy leading to the social optimum steady state in which both agents cooperate. Further, given the interpretability of the AIF, we will be able to analytically derive the process that the agents undergo during this learning process and thus predict how it might change with different parameters.

\section{Simulation Dynamics}\label{sec:sim_dynamics}

\begin{table}[t!]
    \centering
    {\renewcommand{\arraystretch}{1.2}
    \begin{tabular}{ | p{6cm} | p{8cm} | }
        \hline
        \textbf{Variable Name} & \textbf{Notation} \\
        \hline\hline
        Hidden States & $\mathbf{s} \in \{ \mathrm{CC}, \mathrm{CD}, \mathrm{DC}, \mathrm{DD} \}$ \\ \hline
        Observations & $\mathbf{o} \in \{ \mathrm{CC}, \mathrm{CD}, \mathrm{DC}, \mathrm{DD} \}$ \\ \hline
        Actions & $\mathbf{u} \in \{u^{C}, u^{D}\}$ \\ \hline
        Observation Model & $P(\textbf{o}_t | \textbf{s}_t ; A) = Cat(\mathbf{A})$ \\ \hline
        Transition Model & $P(\mathbf{s}_{t+1} | \mathbf{s}_{t-1}, \mathbf{u}_{t-1};B) = Cat(\mathbf{B})$ \\ \hline
        Transition Model Parameter & $P(B) = \prod_{ju} P(B_{\bullet ju}), \hspace{3mm} P(B_{\bullet ju}) = Dir(\mathbf{b}_{\bullet ju})$ \\ \hline
        Initial State Prior & $P(\mathbf{s}_1; D) = Cat(\mathbf{D})$ \\ \hline
        `Biased' State Prior (Reward) & $\tilde{P}(\mathbf{s}; C) = Cat(\mathbf{C}), \hspace{2.0mm} \textrm{s.t.} \hspace{1.0mm} \ln \mathbf{C} = [3,1,4,2]$ \\ \hline

    \end{tabular}}
    \caption{Generative model variables and notation.}
    \label{table:gen_mod}
\end{table}

\begin{figure}[t!]
    \centering
    \includegraphics[width=0.95\textwidth]{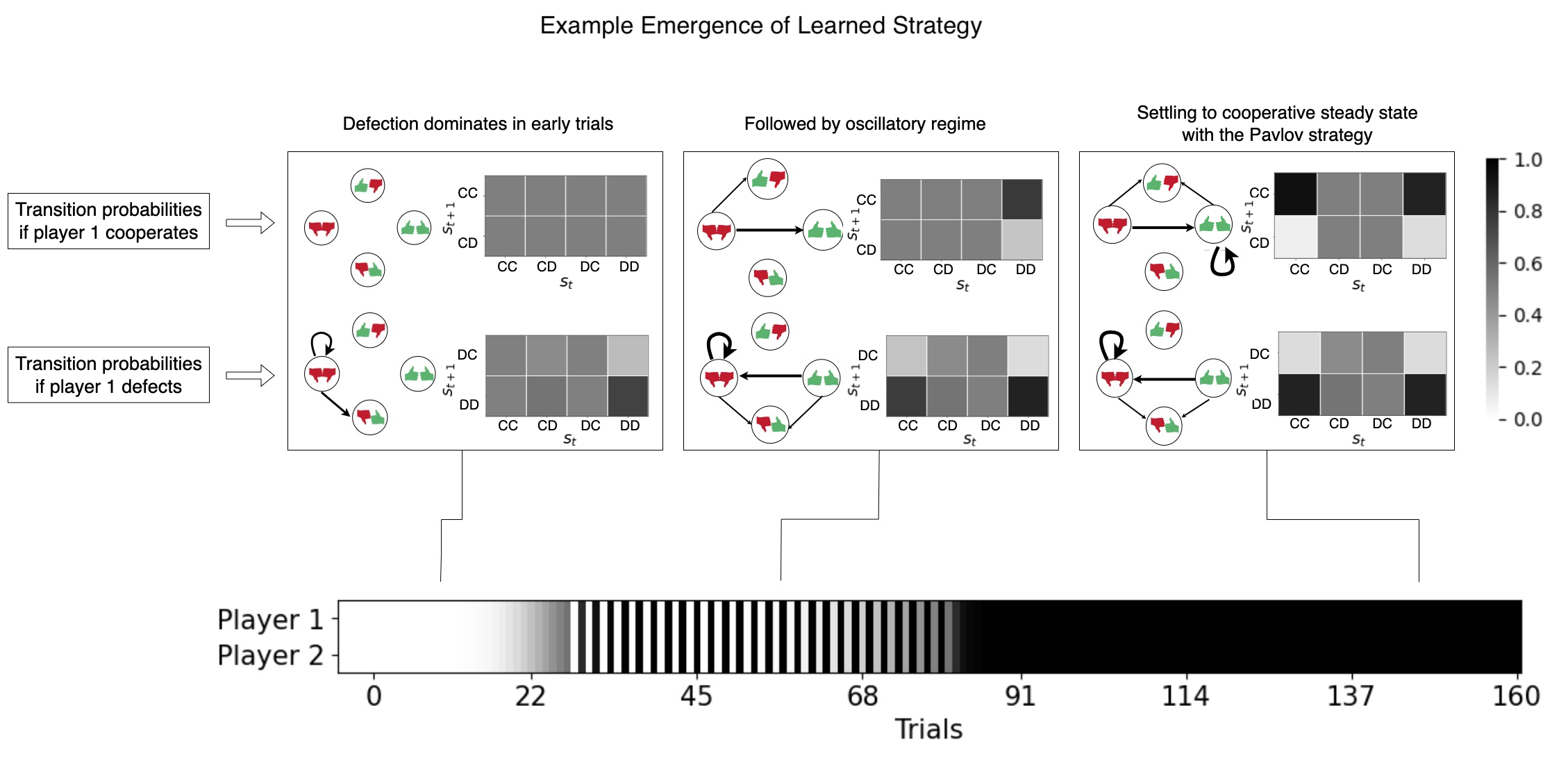}
    \caption{\textbf{Beliefs about transition probabilities over trials.} \textbf{Top:} A representation of Player 1's beliefs at three phases of the simulation $(t=10,20,150)$. Each box contains a graph representation of the transition probabilities, and histograms of the cooperate-conditioned (top row) or defection-conditioned (bottom row) transition distributions at the displayed trial indices. Darker values represent a higher probability. \textbf{Bottom:} The inferred probabilities of cooperation in each trial. Agents select the action with the highest posterior probability. The agents begin by continuously defecting, then undergo an oscillatory period of defection and cooperation, and eventually reach a cooperative steady state. After this period of training, they will have learned the Pavlov strategy, i.e.~they will cooperate if the agent's and opponent's moves are the same in the previous round and defect otherwise \cite{pavlov}.}
    \label{fig:fig1}
\end{figure}

Here, we explore the long-term dynamics of the IPD. Agents play in turns for a finite set of trials, updating their transition model beliefs $Q(B; \boldsymbol{\phi}_{\mathbf{b}})$ at each trial. Unless otherwise specified, agents are configured exactly the same (same priors, same learning rate) and sample their actions deterministically as described in \eqref{eq:deterministic_sampling}. In this model, agents always converge to the cooperative steady state and remain there indefinitely. The magnitude of the learning rate $\eta$ affects the rate of convergence by scaling the update to the transition matrix at each timestep, as shown in \eqref{eq:updateequation}. In Figure \ref{fig:fig1} we show the simulation dynamics for agents configured with learning rate $\eta = 0.3$, but it's important to note that at different learning rates, the nature of these dynamics would not change - rather the critical time points would only occur either sooner (for larger $\eta$) or later (for smaller $\eta$). Therefore, the amount of time taken in order to converge is not representative of the performance of this model, but rather a parameter that can be tweaked. Given the transparency of this deterministic system, it is possible to explain exactly how these agents are `thinking', given their posteriors over time.

Agents are initialised with uniform transition matrices as in \eqref{eq:initialB}. Upon the first observation, they infer the game state and calculate the expected free energies (EFEs, or $\mathbf{G}$) of cooperating and defecting. They take the action that has smaller EFE, i.e., $\arg \min_{u} \mathbf{G}_{0}(u)$. At first, because of the reward parameterization and the uniformity in the transition prior $P(B;\mathbf{b})$, defection will minimise the EFE (i.e., predicts the highest reward), according to:

\begin{align}
    \mathbf{G}_0(u = \mathrm{C}, \phi^{\mathrm{C}}) = -(\mathbf{B}_0^{\mathrm{C}} \cdot \phi_{0}^{\mathrm{C}}) \cdot (\ln \mathbf{B}_0^{\mathrm{C}} \cdot \phi_0^{\mathrm{C}} - \ln \mathbf{C} ) &= \frac{1}{2} \ln (\mathbf{C}_1\mathbf{C}_2) - \ln \frac{1}{2} \\
    \mathbf{G}_0(u = \mathrm{D}, \phi^{\mathrm{D}}) = - (\mathbf{B}_0^{\mathrm{D}} \cdot \phi_0^{\mathrm{D}}) \cdot (\ln \mathbf{B}_0^{\mathrm{D}} \cdot \phi_0^{\mathrm{D}} - \ln \mathbf{C} ) &= \frac{1}{2} \ln (\mathbf{C}_3\mathbf{C}_4) - \ln \frac{1}{2}
    \label{eq:initialG}
\end{align}

Therefore, as long as $\ln (\mathbf{C}_3 \mathbf{C}_4) < \ln (\mathbf{C}_1 \mathbf{C}_2) $, the agent always defects on the first timestep. Agents will then continue to defect, because the expected reward from realising the state DC still outweighs that of any other predicted state. As the agents continue to defect, their beliefs about $P(\mathbf{s}_{t} = \mathrm{DC} | s_{t-1} = \mathrm{DD}, u = \mathrm{D})$ will be decreasing with a proportional increase in $P(\mathbf{s}_{t} = \mathrm{DD} | \mathbf{s}_{t-1} = \mathrm{DD}, u = \mathrm{D})$, meaning $\mathbf{G}(u = D)$ will increase as the probability of getting their desired reward decreases.

At a critical time, which we denote $\tau_1$\footnote{whose solution in terms of generative model parameters we derive in the next section.}, the agents will begin assigning more probability to cooperation than defection $\phi^{C} > \phi^{D}$, because the transition probabilities have decreased sufficiently for the EFE of cooperation to outweigh that of defection. Once the agents begin cooperating, they undergo an oscillatory period during which their actions fluctuate from cooperation to defection. This is because at $\tau_1$, the transition probabilities $P(\mathbf{s}_t | s_{t-1} = \mathrm{CC})$ are fixed at their initial value, since the agents have yet observed the previous state being CC. Thus the agents will still be optimistic about realising the highest reward state DC via the transition probability $P(\mathbf{s}_{t+1} = \mathrm{DC} | \mathbf{s}_t = \mathrm{CC}, u = \mathrm{D})$.

The agents will eventually learn that inferring to defect will inevitably lead to observing DD, and inferring to cooperate will inevitably lead to CC. The oscillatory period is crucial to this because it teaches the agent that defecting in response to cooperation will only ever lead to DD. The oscillation continues until the critical time point $\tau_2$, in which the probability $p(\mathbf{s}_{t+1} = \mathrm{DC} | \mathbf{s}_{t} = \mathrm{CC}, \mathbf{u}_t = \mathrm{D})$ becomes smaller than $p(\mathbf{s}_{t+1} = \mathrm{DD} | \mathbf{s}_{t} = \mathrm{CC}, \mathbf{u}_t = \mathrm{D})$, at which point the agents will cooperate for all remaining rounds.

\subsection{The analytic transition function}

In the above model of AIF agents, an analytic solution for the evolution of each agent's beliefs about the transition likelihood $Q(B; \phi^{*}_{\mathbf{b}})$ is available. This is formulated by deriving approximations to $\tau_1$---the critical trial in which the agents transition to an oscillatory period between defection and cooperation---and $\tau_2$, the second phase transition in which the agents converge to the cooperative steady state. Given the expressions for $\tau_1$ and $\tau_2$ in \eqref{eq:taus}, we can write down the evolution of the Dirichlet parameters of the transition probability matrix. The derivations for the following expressions are in the Appendix (\ref{sec:appendix_tau1}, \ref{sec:appendix_tau2}). Here, $\mathbf{C}$ corresponds to the `biased' state reward prior, and each entry of $\mathbf{C}$ corresponds to the reward value of that observation $(r_{\mathrm{CC}}, r_{\mathrm{CD}}, r_{\mathrm{DC}}, r_{\mathrm{DD}})$. For the full definition see \eqref{eq:reward}.

\begin{equation}
\tau_1 \approx \frac{R_1(\beta)} {\eta}\hspace{2cm} \tau_2 \approx \frac{R_2(\beta)}{\eta}
\label{eq:taus}
\end{equation}
where
\begin{equation}
R_1 = \frac{2}{\ln \frac{\mathbf{C}_3}{\mathbf{C}_4} + 2 - \sqrt{(\ln \frac{\mathbf{C}_4}{\mathbf{C}_3}-2)^2 - 8(-\ln \frac{\mathbf{C}_4}{2\sqrt{\mathbf{C}_1 \mathbf{C}_2}} - \frac{1}{5})}} -1 \hspace{1cm} R_2 = \frac{3}{2}R_1
\end{equation}

This means that $\tau_2$, the number of trials it takes the system to reach the steady state, can be precisely approximated as a linear function of $\tau_1$, the number of trials it takes to start the oscillatory period (see Figure \ref{fig:approximations})---i.e.~the critical time points governing the phase transition are linearly related to each other as a function of learning rate and the reward function. Given these expressions, our analytic form of the transition rule for the posterior Dirichlet parameters over the transition model is:

\begin{equation}
\mathbf{\phi}_{\mathbf{b}_{t+1}}^{\mathbf{C}} =
\begin{cases}
  \mathbf{b}_0^{\mathbf{C}} \\
      \mathbf{b}^{\mathbf{C}}_{0} + \frac{\eta}{2} s^{\mathbf{CC}} \otimes s^{\mathbf{DD}} (t - \frac{R_1 }{\eta} ) \\
      \mathbf{b}^{\mathbf{C}}_{\tau_2} + \eta s^{\mathbf{CC}} \otimes s^{\mathbf{CC}} (t - \frac{R_2}{\eta}) \\
    \end{cases}
    \mathbf{\phi}_{\mathbf{b}_{t+1}}^{\mathbf{D}} =
    \begin{cases}
        \mathbf{b}^{\mathbf{D}}_{0} + \eta s^{\mathbf{DD}} \otimes s^{\mathbf{DD}}t & \vline \hspace{0.2cm} t < \frac{R_1}{\eta} \\
        \mathbf{b}^{\mathbf{D}}_{\tau_1} + \frac{\eta}{2} s^{\mathbf{DD}} \otimes s^{\mathbf{CC}} (t - \frac{R_1 }{\eta} ) & \vline \hspace{0.2cm}\frac{R_1 }{\eta} < t < \frac{R_2}{\eta} \\
        \mathbf{b}^{\mathbf{D}}_{\tau_2} & \vline \hspace{0.2cm}t > \frac{R_2}{\eta} \\
    \end{cases}
    \label{eq:dynamics2}
\end{equation}
which can be used to exactly replicate the trajectory of $Q(s_{t+1} | s_t, u_t)$ over time (Figure \ref{fig:approximations}).

\begin{figure}[t!]
    \centering
    \includegraphics[width=1.0\textwidth]{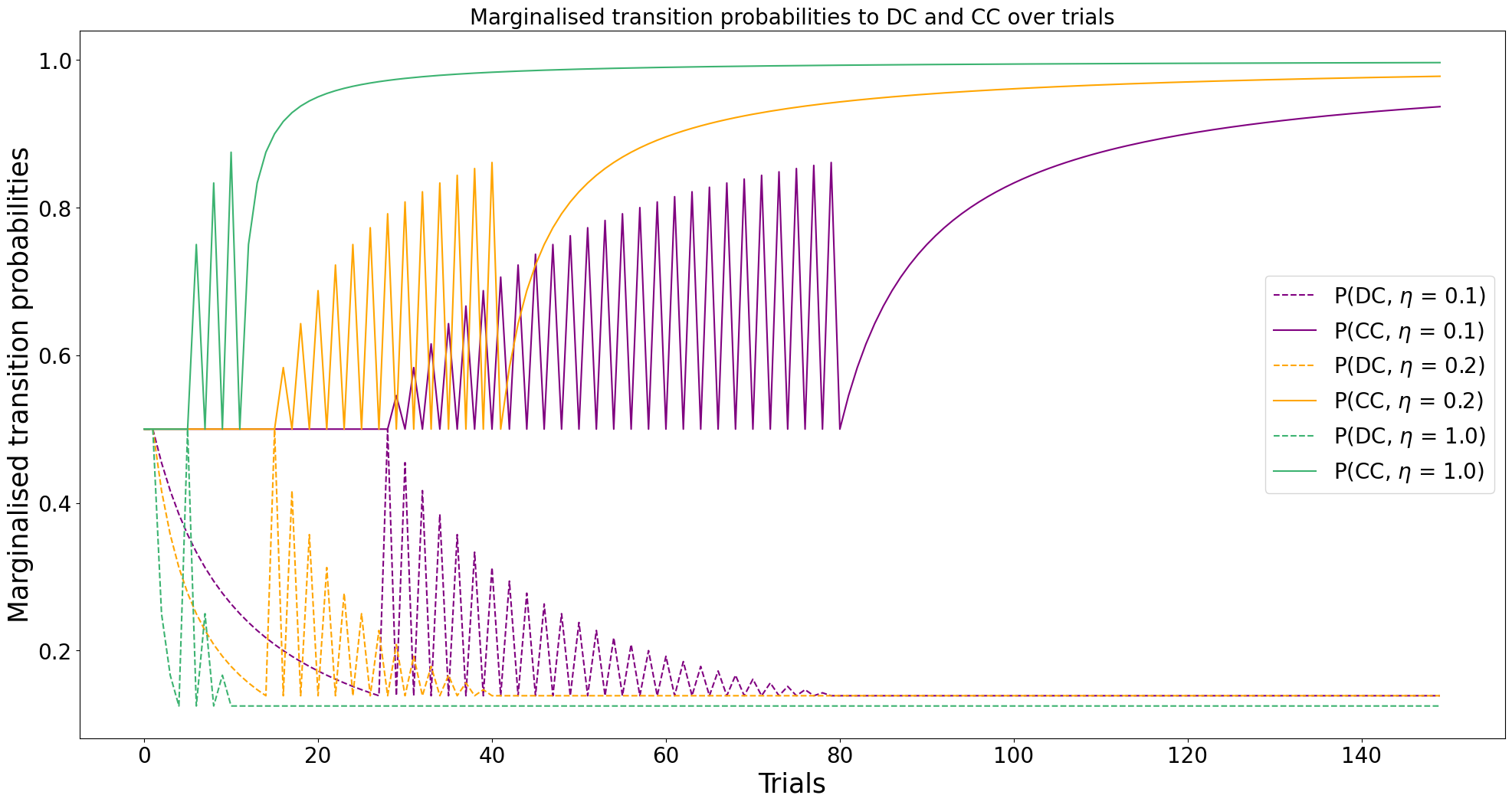}
    \caption{\textbf{Marginalised transition probabilities under different $\eta$.} The dotted lines represent the marginalised probabilities from all states to the highest reward state DC, and the solid lines represent the marginalised probabilities from all states to the socially optimal state CC. The transition probabilities to DC decrease initially during the period of defection, then fluctuate during the period of oscillation and steady out close to 0 once the agents reach the cooperate steady state, and the probabilities to state CC take the same pattern in the opposite direction. This happens more rapidly for larger $\eta$, because the updates to the parameters of the transition likelihood distribution are larger at every trial.}
    \label{fig:actionsovertime}
\end{figure}

We conclude by noting that the agents in this model, after undergoing these two phase transitions and converging to CC, have learned the well-known Pavlov (also known as the ``Win-Stay Lose-Shift'') strategy from IPD literature \cite{pavlov}. Agents learned during $0<t<\tau_1$ that given the observation DD, the best strategy is to cooperate, and during $\tau_1<t<\tau_2$ they learned that cooperating is the best outcome given the observation CC---therefore, having reached $\tau_2$, they continue cooperating. To show that the agents learned the Pavlov strategy, we performed an experiment where once an agent converged to the steady state, we disabled additional learning and had this agent play against an agent that behaves completely randomly. When playing against this random agent, they observe the new asymmetric states DC or CD. The desire to maximise expected utility (via the drive to minimise KL risk, a.k.a., the expected free energy) will lead them to perform the `greedy' strategy of defection, which is how their behaviour is consistent with the Pavlov strategy\footnote{An agent exhibiting the Pavlov strategy will only cooperate if in the previous trial, both agents performed the same action (i.e., the state was either CC or DD, otherwise they will defect).}. Future work will further characterise the space of learnable strategies under this framework.

\section{Generalizing the model}

\begin{figure}[t!]
    \centering
    \includegraphics[width=1.0\textwidth]{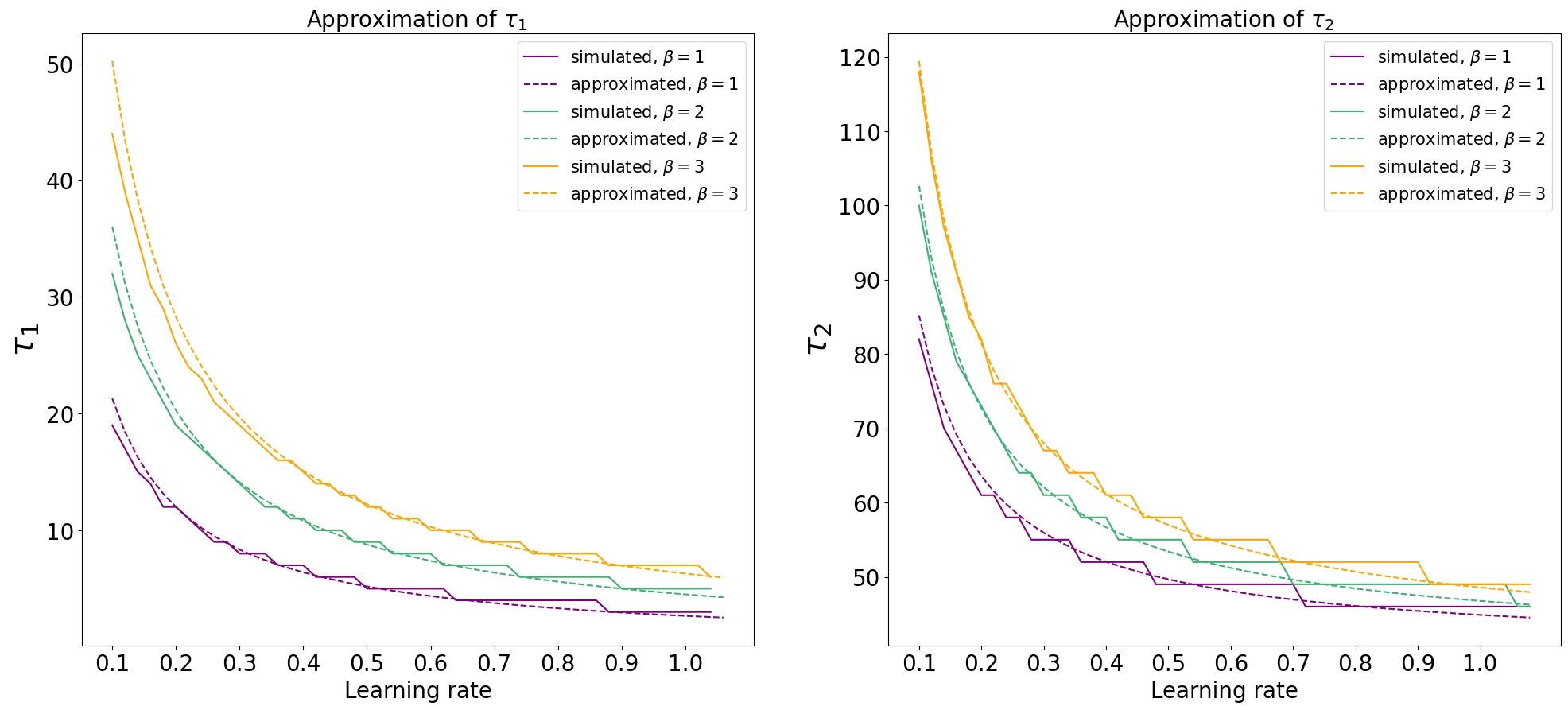}
    \caption{\textbf{Simulated vs.~derived relation between reward and learning rate.} Simulated and approximated $\tau$s for three values of $\beta$ parameterizing the reward function. On the left, we approximate $\tau_1$ with the equation $\tau_1 = \frac{R_1}{\eta}$ where $R_1$ depends on the reward parameter of $\beta$. On the right, we approximate $\tau_2$ with $\tau_2 = \tau_1 + \frac{1}{\eta} R_2$ where again, $R_2$ depends on $\beta$. With a larger $\beta$, meaning a higher predicted reward for the state DC, the values of $\tau$ increase as it will take more trials for the players to update their transition probabilities away from having a preference to defect.}
    \label{fig:approximations}
\end{figure}

In the previous section, we found an approximate solution for the belief-, action-, and learning-dynamics, which completely describes the case of two symmetrically-parameterised agents playing IPD. For any given parameterisation of the prior preferences $\mathbf{C}$, we derived the trials at which the critical transitions take place in the two-agent system, steering it away from the Nash equilibrium and towards the cooperative steady state.

The simplicity of this model is that these agents are configured exactly alike, and therefore there is complete symmetry in the state space. This means that the agents will only ever observe two out of four possible states in the space. However, this case no longer holds when either the agents are parameterised with different learning rates, or when they sample their actions stochastically, according to \eqref{eq:stochastic_sampling}. These cases open the space of possible strategies that the agents can learn, some of which will lead the agents to fall into the Nash equilibrium, and others which will allow them to reach the optimal outcome.

\subsection{Different learning rates}

We now assume agents parameterised with different $\eta$ and the same $\beta$, performing actions deterministically. We denote the agent with larger $\eta_1$ as $a_1$, and the agent with smaller $\eta_2$ as $a_2$. According to \eqref{eq:tau1}, the critical value $\tau_1$ depends on $\eta$, and since $\eta_1> \eta_2$, this means $\tau_1^{a_1} < \tau_2^{a_2}$. Thus, $a_1$ will cooperate at $\tau_1^{a_1} = \frac{R_1}{\eta_1}$, but $a_2$ will not yet deem cooperation a better policy than defection (namely, the EFE of defection will remain below that of cooperation). Therefore, at $\tau_1^{a_1}$, the game state will be CD from $a_1$'s perspective and DC from $a_2$'s perspective. This symmetry-breaking means that the system will not enter into the typical oscillation phase triggered by mutual cooperation (as is guaranteed when $\eta_1 = \eta_2$ and thus $\tau_1^{a_1} = \tau_1^{a_2}$).

The nonidentical observations imply that after $\tau_1^{a_1}$, $a_1$ believes $P(\mathbf{s}_{t+1} = \mathrm{CD} | \mathrm{DD})$ is more probable, thereby being disincentivised to continue cooperating, and $a_2$ believes $P(\mathbf{s}_{t+1} = \mathrm{DC} | \mathrm{DD})$ is more probable, being incentivised to continue defecting. The degree of disincentivisation (or incentivisation) will increase in proportion to $\eta_1$ or $\eta_2$, respectively, due to a corresponding $\eta_1$-scaled increase in $\mathbf{G}^{a_1}(\mathbf{u} = C)$ and an $\eta_2$-scaled decrease in $\mathbf{G}^{a_2}(\mathbf{u} = D)$. This growing asymmetry in the agents' beliefs means that \eqref{eq:dynamics2} no longer holds. At this point, the agents will return to continuous defection until another instance of $\mathbf{G}(\mathbf{u} = D) = \mathbf{G}(\mathbf{u} = C)$ occurs; the duration of this depends on $\eta$.

\begin{figure}[t!]
    \centering
    \includegraphics[width=1.0\textwidth]{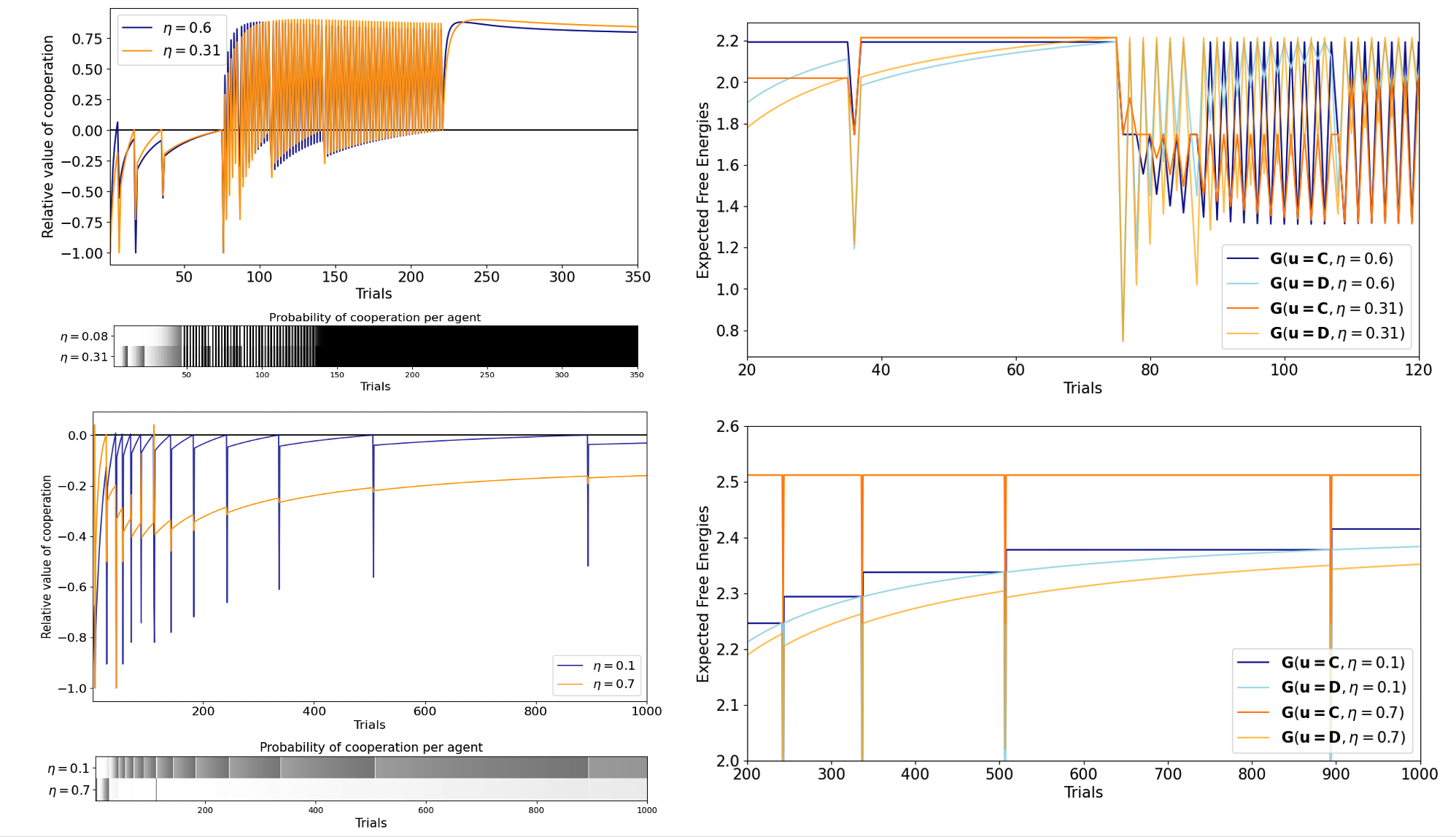}
    \caption{\textbf{Relative value of cooperation under different $\eta$ parameterisations.} \textbf{Above:} Agents are configured with $\eta$s along the tendrils of Figure \ref{fig:heatmaps}. On the left, the relative values of cooperation, calculated as $\mathbf{G}(\mathbf{u}=C) -\mathbf{G}(\mathbf{u}=D)$, reach zero several times and converging around 0.75 at the optimal outcome. On the right: the fluctuations in the individual EFEs. There are periods before $\tau_1$ and between $\tau_1$ and $\tau_2$ in which one player will cooperate and the opponent defects; this creates the spikes in the distribution, as one agent is punished and the other is rewarded. \textbf{Below:} Agents with $\eta$s that are not on the tendrils in Figure \ref{fig:heatmaps}, meaning that they do not converge to the cooperative steady state. We can see on the left how $ \mathbf{G}_t(\mathbf{u}=D)$ is converging to something less than $\mathbf{G}_t(\mathbf{u}=C)$.}
    \label{fig:examples_from_heatmaps}
\end{figure}

In sum, the conditions under which the joint-agent system converges to the optimal steady state is determined by whether or not the agents' learning rates are configured such that there will be some time point $t$ less than some threshold $T_{max}$ in which both agents cooperate simultaneously. If this is not the case, then as defection continues, the rate of increase of $\mathbf{G}(\mathbf{u}=D)$ slows, and after a certain amount of time (governed by $\eta$) it will become too slow and never catch up to $\mathbf{G}(\mathbf{u}=C)$ (see Figure \ref{fig:examples_from_heatmaps}). In other words, if at any point, for either agent, $\mathbf{G}_t(\mathbf{u}=D) < \mathbf{G}_t(\mathbf{u}=C) \hspace{2mm} \forall t \in (0, T_{max}]$, the agents are trapped in the Nash equilibrium.

\begin{figure}[t!]
    \centering
    \includegraphics[width=1.0\textwidth]{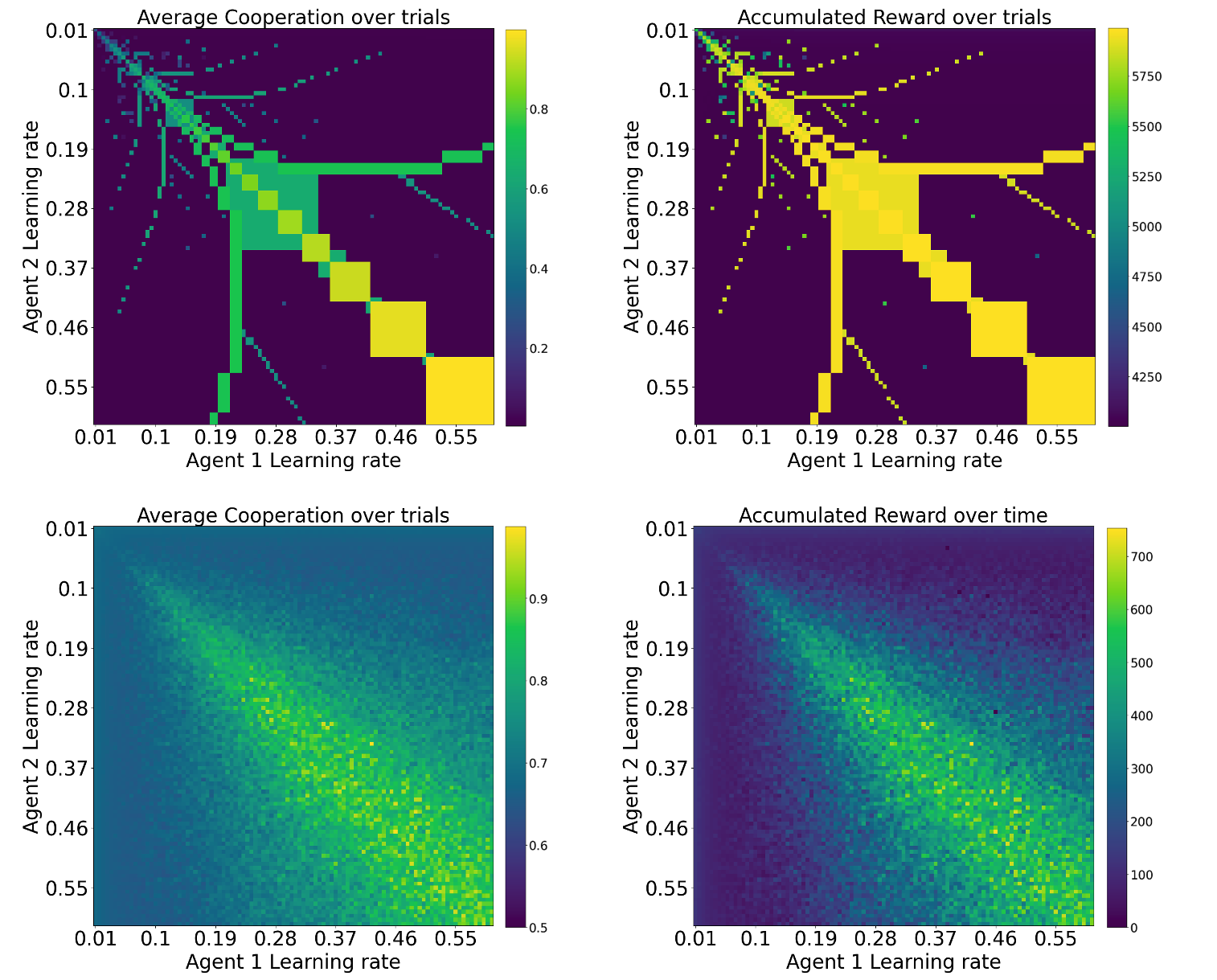}
    \caption{\textbf{Parameter sweeps over $\eta$.} \textbf{Top row:} Agents sample actions deterministically. Wherever the average cooperation is nonzero, agents converged to the cooperative steady state---yellow cells indicate faster cooperation, which is generally associated with higher overall reward. \textbf{Bottom row:} Agents sample actions stochastically. Cooperation still occurs most often along the diagonal, tapering off as learning rates become more different.}
    \label{fig:heatmaps}
\end{figure}

Figure \ref{fig:examples_from_heatmaps} shows EFE trajectories in scenarios where agents converge to the optimal outcome (above) and where agents get trapped in the Nash equilibrium (below). Convergence to the Nash equilibrium occurs in the absence of any trial where the relative value of cooperation reaches 0 simultaneously for both agents. Instead, the relative values of cooperation slowly converge to different and nonoverlapping limits\footnote{Note that even after the agents reach a cooperative steady state, the difference in expected free energy takes time to flatten because the entropy is still decreasing as beliefs become more precise, via learning.}. If the intersection of the condition in \eqref{eq:oscillationsbegin} does occur, this guarantees that the agents will begin the oscillatory period which will eventually lead them to convergence to CC (while there may be some instances of CD and DC in the oscillatory period, this will not prevent eventual cooperation). In general, when learning rates are close together, the likelihood of convergence to CC is more likely; however, the actual pattern is more complicated than this. Figure \ref{fig:heatmaps} demonstrates the complex pattern of instances in which the agents converge to the cooperative steady state given different learning rate combinations, with both the deterministic and stochastic sampling.

\subsection{Stochastic sampling}

Here, we introduce noise in the action selection such that agents sample actions with some probability proportional to their (negative) EFE. Action stochasticity can be controlled with an inverse temperature parameter $\alpha$ according to \eqref{eq:stochastic_sampling}. In general, all of the principles outlined in Section \ref{sec:sim_dynamics} remain; however, now the agents will sometimes perform the suboptimal action. This enables agents to experience the entire state space (different combinations of defection and cooperation) and therefore estimate transitions between all the combinations of states.

We can see from Figure \ref{fig:heatmaps} that, on average, endowing the agents with stochasticity enables them to converge to the cooperative steady state for a larger number of combinations of different learning rates. This makes sense, because it increases the likelihood of `escaping' the pattern of continuous defection, and therefore learning about the advantages of cooperation. In terms of the reward, the agents that have most similar learning rates will behave most similarly and therefore accumulate more reward (along the diagonal).

\section{Conclusion}

Iterated Prisoners' Dilemma games have long been the test bed for new developments in behavioural science and game theory. Because of the relative simplicity of the game's structure---and its, at times, surprising experimental results---researchers often use it to develop mathematical frameworks for understanding decision making in social or multi-agent contexts. In this paper, we demonstrated how active inference can be used to model the IPD transparently, such that in a simple set-up, we can derive a solution to the evolution of the agents' beliefs about the game dynamics, i.e., the transition probabilities. This allows us to quantitatively reason about why the agents converge to their chosen optimal strategy and how behaviour changes as a function of different learning rates and stochastic action selection. While the simple case of similarly-configured agents resulted in both agents exhibiting the Pavlov strategy, once we introduce asymmetry in the generative models, and/or stochasticity in action sampling, then upon testing, agents are able to learn a variety of different strategies, including the Pavlov strategy, Unconditional Defection, Unconditional Cooperation, and Tit for Tat---or some variation of Tit for Tat \cite{imhof2007tit, wedekind1996human}.

This finding is a starting point for future work, in which such a model could be extended to multiple agents interacting towards a common goal, and investigating the various strategies that emerge from acting in a network order to minimise free energy. The current model did not incorporate the information-seeking components that are often leveraged in action-selection under active inference \cite{friston2015active}. In our case, the ambiguity term of the expected free energy was zero by construction (due to zero observation uncertainty), but future work could explore the role of parameter information gain (resolving uncertainty about $B$) and how that changes the multi-agent dynamics in IPD. Overall, in this work we demonstrated that AIF can offer game theory a novel analytic transparency and simplicity for accounting for multi-agent dynamics using a first-principles, Bayesian account.

\paragraph{Acknowledgements:} The authors thank Wolfram Barfuss and Christoph Riedl for valuable feedback and comments that substantially improved the quality of the manuscript. \textbf{Funding information: } DD, CH, \& BK acknowledge the support of a grant from the John Templeton Foundation (61780). The opinions expressed in this publication are those of the authors and do not necessarily reflect the views of the John Templeton Foundation.

\begin{sloppypar}
\printbibliography[title={References}]
\end{sloppypar}

\clearpage
\appendix
\setcounter{figure}{0}
\setcounter{table}{0}
\setcounter{equation}{0}
\renewcommand\thefigure{\thesection.\arabic{figure}}
\renewcommand\thetable{\thesection.\arabic{table}}
\renewcommand\theequation{\thesection .\arabic{equation}}
\begin{refsection}

\section{Supplementary Information}\label{sec:appendix_A}

\subsection{Generative Model}

In this section, we describe the Prisoner's Dilemma game as a two-agent active inference system and determine the conditions under which the agents reach the optimal state of constant cooperative play, avoiding the Nash equilibrium. To enable active inference agents to reach the cooperative steady state, we invoke the notion of parameter learning; specifically, the ability of agents to infer likely sequences of game states by updating posterior beliefs about transition probabilities. These transition probabilities parameterise a likelihood model that describes transitions between game states (e.g., the transition from the state of `cooperate-cooperate' to `cooperate-defect'). Under active inference, this parameter learning is cast as a problem of inferring generative model parameters. Usually, parameter inference unfolds on a slow timescale (hence the term `learning') relative to `fast' inference of hidden states \cite{friston2016active} (See Table \ref{table:gen_mod_si} for full description of model parameters).

\begin{table}[h]
    \centering
    {\renewcommand{\arraystretch}{1.3}
    \begin{tabular}{ | p{6.5cm} | p{8.5cm} | }
        \hline
        \textbf{Variable Name} & \textbf{Notation} \\
        \hline\hline

        Hidden States & $\mathbf{s} \in \{ \mathrm{CC}, \mathrm{CD}, \mathrm{DC}, \mathrm{DD} \}$ \\ \hline

        Observations & $\mathbf{o} \in \{ \mathrm{CC}, \mathrm{CD}, \mathrm{DC}, \mathrm{DD} \}$ \\ \hline

        Actions & $\mathbf{u} \in \{u^{C}, u^{D}\}$ \\ \hline

        Observation Model & $P(\textbf{o}_t | \textbf{s}_t ; A) = Cat(\mathbf{A})$ \\ \hline

        Transition Model & $P(\mathbf{s}_{t+1} | \mathbf{s}_{t-1}, \mathbf{u}_{t-1};B) = Cat(\mathbf{B})$ \\ \hline

        Transition Model Parameter & $P(B) = \prod_{ju} P(B_{\bullet ju}), \hspace{3mm} P(B_{\bullet ju}) = Dir(\mathbf{b}_{\bullet ju})$ \\ \hline

        Initial State Prior & $P(\mathbf{s}_1; D) = Cat(\mathbf{D})$ \\ \hline

        `Biased' State Prior (Reward) & $\tilde{P}(\mathbf{s}; C) = Cat(\mathbf{C}), \hspace{2.0mm} \textrm{s.t.} \hspace{1.0mm} \ln \mathbf{C} = [3,1,4,2]$ \\ \hline

    \end{tabular}}
    \caption{Generative model variables and notation.}
    \label{table:gen_mod_si}
\end{table}

The agent's generative model is a Markov Decision Process \cite{puterman1990markov} that encodes a joint distribution over sequences of hidden states $\mathbf{s}_{1:T}$ observations $\mathbf{o}_{1:T}$, actions $\mathbf{u}_{1:T}$, and model parameters $A, B, D$ \cite{Pymdp2022}. Markov Decision Processes assume that the dynamics are shallow, with single-timestep dependency $P(s_{t+1}|s_t, u_t; B)$; this Markov property means we can write the generative model as a product of time-dependent distributions:

\begin{align}
    P(\mathbf{o}_{1:T}, \mathbf{s}_{1:T}, \mathbf{u}_{1:T}, A, B) &= P(\mathbf{s}_1; D)P(\mathbf{\pi})P(A)P(B)P(D)\prod_{t=1}^{T-1} P(\mathbf{o}_{t+1} | \mathbf{s}_{t+1}; A)P(\mathbf{s}_{t+1} | \mathbf{s}_{t},\mathbf{u}_{t}; B)
\end{align}
multiplied by initial priors over hidden states, policies, and parameters.

The hidden states $\mathbf{s}$ consist of a single factor with four possible states or levels, corresponding to the game states (the four combinations of possible two-player choices): CC, CD, DC, and DD. This game state factor comprises the primary random variable in each agent's model.

In our notation, the first letter of each game state corresponds to the focal agent's choice, and the second letter corresponds to that of its opponent. In our formulation, agents have precise knowledge of the current game state, which they technically infer through (unambiguous) observation of their and their opponent's action. Uncertainty comes into the game insofar as agents must \textit{predict} the subsequent game state and then act based on their predictions and their desires to maximise utility.

There is one observation modality with four observations, which again correspond directly to the four game states. Therefore, the four observations are CC, CD, DC, and DD. Note that the agents will only observe the game state after-the-fact, i.e., each observation corresponds to the game state in the previous round of iterative play. This is because in the Prisoner's Dilemma, the agents perform their actions at any given trial without knowing what their opponent will do in that trial, but in iterative play, the agents can build a strategy over time by observing the resulting game states after each trial ends.

\subsubsection{Observation Likelihood}

The observation model $P(\mathbf{o}_t|\mathbf{s}_t, A)$ is a conditional distribution encoding the agent's beliefs about the relationship between the current (hidden) game state and its concurrent observation. Also known as the likelihood model, the agent uses this distribution to infer the most likely game state, given an observation thereof.

In the simulations presented here, we assume that agents are equipped with a deterministic, unambiguous observation model, i.e., observations are deterministic indicators of the game state. In the discrete state space models common in active inference, likelihoods like $P(\mathbf{o}_t|\mathbf{s}_t, A)$ are often represented as multidimensional arrays (e.g., matrices) whose values are populated by parameters; in the case of the observation model, we represent this likelihood directly as a matrix $\mathbf{A}$ whose entries are given by the likelihood parameters $A$. Hereafter we use boldface $\mathbf{X}$ to indicate a representation of Categorical parameters in terms of vectors and matrices, and use the standard italic notation $X$ to indicate the random variable in the generative model (e.g. $P(A)$). When we have an unambiguous or precise likelihood mapping, this matrix is the identity matrix, representing the mapping from hidden states (columns) to observations (rows):

\begin{align}
    P(\mathbf{o}_t = i | \mathbf{s}_t =j, [A]_{ij}) = \delta(i-j) \label{eq:obs_lik1}\\\notag\\
    \mathbf{A} = \begin{bmatrix}
        1 & 0 & 0 & 0 \\
        0 & 1 & 0 & 0 \\
        0 & 0 & 1 & 0 \\
        0 & 0 & 0 & 1
    \end{bmatrix} = \mathbf{I} \label{eq:obs_lik2}
\end{align}

An agent with such a precise likelihood model will infer the game state in the previous round of iterative play entirely based on the observed game state.

However, one can imagine introducing uncertainty into an agent's beliefs by adding off-diagonal, positive values into the $A$ matrix -- this would correspond to the agent believing that game state observations are ambiguous with respect to the true game state. Concretely, we could imagine that one agent might receive a misleading signal indicating that its opponent defected when they actually cooperated. A simple way to parameterise this uncertainty is through an inverse temperature parameter $\psi$, which makes the $A$ matrix totally uninformative (maximum entropy columns) in the limit of $\psi \to 0$, and infinitely precise in the limits of $\psi \to \infty$:

\begin{align}
    \mathbf{A} = \dfrac{\mathbf{I}^{\psi}}{\sum \mathbf{I}^{\psi}}
    \label{eq:likelihood}
\end{align}

Finally, it is worth mentioning that we assume $P(A)$ is infinitely precise and not subject to learning. Therefore, we emit any parameterisation of the priors over this likelihood, while we keep them for the transition likelihood parameters $B$, as we will update these in learning.

\subsubsection{Reward}

Different game states are assigned different rewards or desirabilities under the Prisoner's Dilemma problem formulation. Active inference converts the notion of `reward' into prior probability by equipping agents with biased prior beliefs about future states or observations \cite{friston2009reinforcement}. In the context of planning actions, this biased prior serves the role of a ``goal-vector'' or reward function \cite{Pymdp2022}. We denote this as a biased prior over states in our agent's model $\tilde{P}(\mathbf{s}; \mathbf{C})$\footnote{Note that in many formulations of active inference this is formulated as a prior over observations $\tilde{P}(o;\mathbf{C})$.}. This special `goal prior' is parameterised by a vector of Categorical parameters $\mathbf{C}$. Reward and prior probability can be straightforwardly related via the relation $\tilde{P}(\mathbf{s}) \propto \exp(r)$ \cite{millidge2020relationship}; therefore, we typically parameterise $\mathbf{C}$ using relative log probabilities or nats, i.e., $\mathbf{C} = \ln \tilde{P}(\mathbf{s}) + Z$. Following from Table \ref{table:rewards}, the most desirable observation is $s^{\mathbf{DC}}$ (the agent defects and the opponent cooperates), followed by $s^{\mathbf{CC}}$ (both players cooperate), then $s^{\mathbf{DD}}$ (both players defect), and finally $s^{\mathbf{CD}}$ (the agent cooperates and the opponent defects). Therefore, our $\mathbf{C}$ vector is $\mathbf{C} = [3,1,4,2]$.

Note that the values of these numbers have an effect on the desirability of the observations and therefore will impact the agents action-planning such that they plan actions that they infer will result in the observation of the most desirable state. Changing the values of these rewards will change the incentive and behaviour of the agents.

\subsubsection{Different reward parameterizations}

We can parameterise the reward function $\mathbf{C}$ in terms of a single precision that makes a single ordered reward function with the constraints $r_{\mathrm{CD}} < r_{\mathrm{DD}} < r_{\mathrm{CC}} < r_{\mathrm{DC}}$ more or less shallow/steep. We do this using the softmax (normalised exponential transformation):

\begin{align}
    \mathbf{C} = \sigma\left(\begin{bmatrix}r_{\mathrm{CC}} \\ r_{\mathrm{CD}} \\ r_{\mathrm{DC}} \\ r_{\mathrm{DD}} \end{bmatrix}, \beta\right), \;\; \mathrm{where} \;\; \mathbf{C}_{\mathrm{CC}} &= \frac{\exp(\beta r_{\mathrm{CC}})}{\sum_{i} \exp(\beta r_{i}) } \notag \\
    \ln\mathbf{C}_{\mathrm{CC}} &= \beta r_{\mathrm{CC}} - \ln \left(\sum_{i} \exp(\beta r_{i})\right) \notag \\
    \implies \ln\mathbf{C} &\propto \beta \begin{bmatrix}r_{\mathrm{CC}} \\ r_{\mathrm{CD}} \\ r_{\mathrm{DC}} \\ r_{\mathrm{DD}} \end{bmatrix}
    \label{eq:reward}
\end{align}

\subsubsection{Policies}

A policy $\pi$ is comprised of individual actions, or control states, $\pi = \{\mathbf{u}_1, \mathbf{u}_2, ... \mathbf{u}_{H}\}$. At each trial of iterative play, the agents can either defect or cooperate. This means that the policy space consists of two control states, namely $u^{\mathbf{C}}$ and $u^{\mathbf{D}}$. Once the action is inferred, the intersection of both agents' actions will result in the realised game state.

\subsubsection{Transition Likelihood}

The transition matrix encodes the beliefs that the agent holds about how game states will evolve given previous trials and their actions. Because action selection under active inference depends on model-based planning, this transition model also directly determines the agent's strategy. Although in this work we focus on how agents can automatically learn the game's dynamics and thus their strategies through experience, we nevertheless begin by constraining what agents can learn by initialising agents` beliefs about transition dynamics, so that they assume that two game state transitions are always impossible. Agents believe that when they cooperate, there is zero probability that the next state will be DC or DD, and conversely, when they defect, they believe there is zero probability that the next state will be CD or CC. Therefore, the transition matrix encodes the agent's assumptions about whether the other will cooperate or defect in the next trial, given the outcome of the current trial and the agent's own action.

We use existing formulations of parameter learning under active inference to allow our agents to update their beliefs about transition model over time based on experience. Technically, the agents are updating a Dirichlet posterior belief over the Categorical parameters $B$ that characterise its transition model (a transition probability matrix, mapping from past to current game states, further conditioned on action). They update this matrix of posterior Dirichlet parameters at the end of each trial, based on that trial's outcome.

At the beginning of iterative play, the agent will be initialised with no prior opinion or knowledge about which of the possible transitions are more likely given its actions (aside from the zero constraints laid out above). These uniform initial transition distributions are shown in \eqref{eq:initialB1} and \eqref{eq:initialB}.

\begin{align}
    P(\textbf{s}_{t+1}|\textbf{s}_t, \textbf{u}_t = C) = \begin{bmatrix}
    0.5 & 0.5 & 0.5 & 0.5 \\
    0.5 & 0.5 & 0.5 & 0.5\\
    0 & 0 & 0 & 0 \\
    0 & 0 & 0 & 0
    \end{bmatrix}
    \label{eq:initialB1}
\\\notag\\
    P(\textbf{s}_{t+1}|\textbf{s}_t, \textbf{u}_t = D) = \begin{bmatrix}
    0 & 0 & 0 & 0 \\
    0 & 0 & 0 & 0\\
    0.5 & 0.5 & 0.5 & 0.5 \\
    0.5 & 0.5 & 0.5 & 0.5
    \end{bmatrix}
    \label{eq:initialB}
\end{align}

At the conclusion of each trial during a session of iterative play, a given agent observes the game state of the previous trial and updates its beliefs about transitions based on the realised states and its actions. As these transition dynamics are learned, the agent is simultaneously learning a strategy based on planning the most optimal action (cooperate or defect), given its evolving beliefs.

\subsection{Inference}

\subsubsection{State inference}
At each trial of iterative play, the agents first infer the game state by inverting their Markovian (POMDP) generative model using ongoing observations $\mathbf{o}_t$.

The agent's hidden state inference involves optimising a variational posterior over hidden states and policies $Q(\mathbf{s}_{1:T}, \pi)$ as a categorical distribution with parameters $\tilde{\boldsymbol{\phi}}$ that are factorised `mean-field'-style across timesteps
\cite{blei2017variational}:

\begin{align}
    Q(\mathbf{s}_{1:T}, \pi; \tilde{\boldsymbol{\phi}}) &= Q(\pi; \phi_{\pi}) \prod_{1:T} Q(\mathbf{s}_t; \boldsymbol{\phi}_{\mathbf{s}, t}) \notag
\end{align}

Where the variational parameters $\tilde{\boldsymbol{\phi}} = \{\phi_{\pi}, \boldsymbol{\phi}_{\mathbf{s}_{1:T}}\}$ are themselves segregated into policy-specific parameters $\phi_{\pi}$ and hidden-state-specific parameters $\boldsymbol{\phi}_{\mathbf{s}_{1:T}}$.

At each timestep $t$, the agent performs inference by optimising the posterior parameters $\tilde{\boldsymbol{\phi}}$ to minimise the timestep-specific variational free energy $\mathcal{F}_t$, which due to the Markovian factorisation of the generative model and mean-field factorisation of the posterior, can be expressed in terms of only the generative model of the current timestep $P(\mathbf{o}_t, \mathbf{s}_t, \pi, \mathbf{A}, \mathbf{B}, \mathbf{C})$:

\begin{align}
    \mathcal{F}_t &= \mathbb{E}_{Q(\mathbf{s}_t, \pi; \tilde{\boldsymbol{\phi}}_t)}\left[\ln Q(\mathbf{s}_t, \pi; \tilde{\boldsymbol{\phi}}_t) - \ln P(\mathbf{o}_t, \mathbf{s}_t, \pi, \mathbf{A}, \mathbf{B}, \mathbf{C})\right] \label{eq:single_timestep_VFE}
\end{align}

The optimal posterior parameters $\tilde{\boldsymbol{\phi}}^{*}$ are those that minimise the free energy in \eqref{eq:single_timestep_VFE} and can be found by solving exactly for the fixed points of $\mathcal{F}_t$. We begin by solving for the parameters of the variational beliefs about hidden states $\boldsymbol{\phi}_{\mathbf{s}_t}$:

\begin{align}
    \frac{\partial \mathcal{F}_t}{\partial \boldsymbol{\phi}_{\mathbf{s}_t}} &= 0 \notag \\ \implies \boldsymbol{\phi}^{*}_{\mathbf{s}_t} &= \sigma \left( \ln \mathbf{A}^{T}\mathbf{o}_t + \ln (\mathbf{B}_{\mathbf{u}_{t-1}} \cdot \boldsymbol{\phi}^{*}_{\mathbf{s}_{t-1}}) \right) \label{eq:fixed_point_update}
\end{align}
where $\sigma$ represents the softmax (or normalised exponential) transform of a vector. The $i$\textsuperscript{th} entry of the softmaxed output is given by:

\begin{align}
    \sigma(x)_i &\triangleq \frac{\exp(x_i)}{\sum_j \exp(x_j)}
\end{align}

The initial matrix-vector product in the last line of \eqref{eq:fixed_point_update} $\ln \mathbf{A}^{T}\mathbf{o}_t$ represents the contribution of sensory evidence to inference, and can be thought of as picking out the row of the $\mathbf{A}$ matrix that corresponds to the observation at timestep $t$. The second matrix vector product $\ln (\mathbf{B}_{\mathbf{u}_{t-1}} \cdot \boldsymbol{\phi}^{*}_{\mathbf{s}_{t-1}})$ represents the contribution of prior information to inference. This simple form is a consequence of the mean-field factorisation of the variational parameters $\boldsymbol{\phi}_{\mathbf{s}_{1:T}}$ across timesteps and an `empirical prior' assumption, where the prior term of the generative model $P(\mathbf{s}_{t}) = \mathbb{E}_{P(\mathbf{s}_{t-1})}[P(\mathbf{s}_{t}|\mathbf{s}_{t-1},\mathbf{u}_{t-1}, \mathbf{B})]$ is evaluated at the parameters of the previous timestep's variational posterior, in a manner reminiscent of a belief propagation step or empirical Bayes:

\begin{align}
    P(\mathbf{s}_{t}) &= \mathbb{E}_{P(\mathbf{s}_{t-1})}\left[P(\mathbf{s}_{t}| \mathbf{s}_{t-1},\mathbf{u}_{t-1}, \mathbf{B})\right] \notag \\
    &\approx \mathbb{E}_{Q(\mathbf{s}_{t-1};\boldsymbol{\phi}_{\mathbf{s}_{t-1}})}\left[P(\mathbf{s}_{t}| \mathbf{s}_{t-1},\mathbf{u}_{t-1}, \mathbf{B})\right] \label{eq:empirical_prior}
\end{align}

We can simplify the expression for the parameters of the variational beliefs due to the unambiguous form of the observation likelihood with infinite precision in \eqref{eq:obs_lik2}, $\mathbf{A} = \dfrac{\mathbf{I}^{\psi}}{\sum \mathbf{I}^{\psi}}$, as well as the fact that the agents are taking identical actions at every trial, thus limiting the state space to \{CC, DD\} which implies that inference can be solved for exactly for any trial $t > 0$ as

\begin{align}
    \boldsymbol{\phi}^{*}_{\mathbf{s}_t} &= \sigma \left( \ln \left( \dfrac{\mathbf{I}^{\psi}}{\sum \mathbf{I}^{\psi}}\right)^{T}\mathbf{o}_t + \ln \left(\mathbf{B}_{\mathbf{u}_{t-1}} \cdot \boldsymbol{\phi}^{*}_{\mathbf{s}_{t-1}}\right) \right) \\
    &= \lim_{\psi \rightarrow \infty} \sigma\left(\psi \ln \left(\mathbf{I}^T \mathbf{o}_t\right) + \ln\left(\mathbf{B}_{\mathbf{u}_{t-1}}\cdot \boldsymbol{\phi}^{*}_{\mathbf{s}_t}\right) - \ln \sum \mathbf{I}^{\psi}\right)\\
    &= \sigma\left(\ln\left(\mathbf{I}^{T}\mathbf{o}_1\right) \right) = \mathbf{I}^{T}\mathbf{o}_1
\end{align}

\subsubsection{\label{subsec:model_planning}Policy inference}

Under active inference, action selection and planning are cast as an inference problem, where policies are treated as a latent variable to be inferred. This has deep homology to contemporary approaches to model-based planning in reinforcement learning, such as planning as inference and control as inference \cite{attias2003planning, abdolmaleki2018maximum, botvinick2012planning, millidge2020relationship}. In particular, active inference agents optimise a variational posterior over policies $Q(\pi)$. However, because policies inherently require estimation of future, unobserved states, we use an augmented, `predictive' generative model to perform this policy inference. This predictive generative model is importantly augmented with the biased prior distribution over states $\tilde{P}(\mathbf{s};C)$. Beliefs about policies, similar to those about hidden states, are optimised by minimising a free energy functional of beliefs about the consequences of action under the predictive generative model. This functional is known as the \textit{expected free energy} and exhibits many desirable properties such as a natural balance between information-seeking (`exploration') and goal-directedness (`exploitation') \cite{millidge2021whence}. The approximate posterior over policies $Q(\pi)$ is also a Categorical distribution with parameters $\boldsymbol{\phi}_{\mathbf{u}}$; the optimal setting of these parameters $\boldsymbol{\phi}^{*}_{\mathbf{u}}$ minimises the expected free energy, leading to the relationship:

\begin{align}
    Q(\pi;\boldsymbol{\phi}_{\mathbf{u}}) &= \sigma(-\mathbf{G}(\pi)) \notag \\
    \mathbf{G}(\pi) &= \sum_{\tau=1}^{H} \mathbf{G}_{t+\tau}(\mathbf{u}_{t+\tau-1})
\end{align}

The second line shows that the expected free energy of a policy is the sum of the expected free energies that accrue for each action that comprises the policy: $\pi = \{\mathbf{u}_1, \mathbf{u}_2, ... \mathbf{u}_{H}\}$. For the present purposes we only consider 1-step ahead policies ($H = 1$). This means that the expected free energy of a policy is simply the expected free energy computed one timestep into the future $\mathbf{G}_{t+1}(\mathbf{u}_{t})$.

The expected free energy can be decomposed into expected ambiguity and risk terms:

\begin{align}
    \mathbf{G}_{t+1}(\mathbf{u}_t) = \mathbf{E}_{Q(\mathbf{s}_{t+1}|u_t)}\left[\mathbf{H}\left[P(\mathbf{o}_{t+1}|\mathbf{s}_{t+1})\right]\right] + \operatorname{D}_{KL}\left(Q(\mathbf{s}_{t+1}|\mathbf{u}_t) \parallel \ln P(\mathbf{s}_{t+1}|C)\right)
\end{align}

We can write this general expression in terms of sufficient statistics of the variational distribution over hidden states $\boldsymbol{\phi}^{*}_{\mathbf{s}_t}$. The ambiguity term of the expected free energy vanishes because the agent's likelihood matrix is the identity:

\begin{align}
    \mathbf{A} \mathbf{B}_t \cdot \boldsymbol{\phi}^{*}_{\mathbf{s}_t} \cdot \left( \ln (\mathbf{A}\mathbf{B}_t \cdot \boldsymbol{\phi}^{*}_{\mathbf{s}_t}) - \ln \mathbf{C} \right)\\ 
    = \mathbf{B}_t \cdot \boldsymbol{\phi}^{*}_{\mathbf{s}_t} \left( \ln \mathbf{B}_t \cdot \boldsymbol{\phi}^{*}_{\mathbf{s}_t} - \ln \mathbf{C} \right) -\underbrace{(\mathbf{A} \ln\mathbf{A})\cdot \boldsymbol{\phi^{*}_{\mathbf{s}_t}}}_{=0}
    \label{eq:efe}
\end{align}

\subsubsection{\label{subsec:model_action}Action Selection}

Having optimised a posterior over policies (which in this context simply reduce to control states), action selection simply consists of sampling the action at trial $t$ that minimises the expected free energy, i.e., sampling an action from the posterior marginal over actions.

\begin{align}
    \boldsymbol{\phi}_{\mathbf{u}} &= \sigma (-\mathbf{G} ) \\
    u_{t+1} &\sim Q(\mathbf{u}_{t+1}; \boldsymbol{\phi}_{\mathbf{u}})
\end{align}

This can be done either deterministically by selecting the most probable control state at every timestep:

\begin{align}
    u_{t+1} &= \argmax_{u} Q(\mathbf{u}_{t+1}; \boldsymbol{\phi}_{\mathbf{u}})
    \label{eq:deterministic_sampling}
\end{align}

Or, this can be done stochastically by sampling from the posterior over actions. The stochasticity of this sampling can be further tuned by sampling from a transformed action posterior scaled by a temperature parameter $\alpha$.

\begin{align}
    u_{t+1} \sim Q(\mathbf{u}_{t+1}; \boldsymbol{\phi},\alpha )
    \label{eq:stochastic_sampling}
\end{align}

\subsubsection{B matrix learning}
After every trial of iterative play, each agent updates its posterior beliefs about the transition model $B$ by optimizing Dirichlet parameters $\boldsymbol{\phi}_{\mathbf{b}}$, which are the sufficient statistics of a Dirichlet parameterization of the posterior $Q(B;\boldsymbol{\phi}_{\mathbf{b}})$. This is also known as `learning' in the active inference literature, and analogised to neuronal processes such as synaptic plasticity, which typically occurs on a slower timescale than hidden state inference (analogised to rapid dynamics of neural firing rates) \cite{friston2016active}. Dirichlet distributions are used as the parameterizations of discrete Categorical likelihood matrices, due to their natural role as conjugate priors for the Categorical distribution.

We supplement the generative model with an additional prior over the parameters of the transition model, the Dirichlet distribution $P(B;\mathbf{b})$ parameterised by a vector of positive real hyperparameters $\mathbf{b}$, that can also be interpreted as `pseudocounts', i.e., how many times has the agent seen this particular transition occur, before the simulation starts. Alongside this prior we introduce a variational posterior over $B$ that is also a Dirichlet distribution $Q(B;\boldsymbol{\phi}_{\mathbf{b}})$. This leads to a new expression for the variational free energy at a given time point, which includes an additional Kullback-Leibler divergence between the variational and generative model Dirichlet distributions over $B$ \cite{Pymdp2022}:

\begin{align}
    \mathcal{F}_t &= \mathbb{E}_{Q(\mathbf{s}_t, \mathbf{u}_t, B;\tilde{\boldsymbol{\phi}})}\left[\ln Q\left(\mathbf{s}_t, \mathbf{u}_t, B; \tilde{\boldsymbol{\phi}}\right) - \ln P\left(\mathbf{o}_t, \mathbf{s}_t, \mathbf{u}_t, A, B, C; \mathbf{A}, \mathbf{b}, \mathbf{C}\right)\right] \notag \\
    &= \mathbb{E}_{Q\mathbf{s}_t, \mathbf{u}_t; \boldsymbol{\phi}_{\mathbf{s}, \mathbf{u}}}\left[\ln Q\left(\mathbf{s}_t, \mathbf{u}_t; \boldsymbol{\phi}_{\mathbf{s}, \mathbf{u}}\right) - \ln P\left(\mathbf{o}_t, \mathbf{s}_t, \mathbf{u}_t, A, C; \mathbf{A}, \mathbf{C}\right)\right] + \operatorname{D}_{KL}\left(Q(B; \boldsymbol{\phi}_{\mathbf{b}})\parallel P(B; \mathbf{b})\right)
\end{align}

This new expression means that when we minimise $\mathcal{F}_t$ with respect to the variational (Dirichlet) parameters $\boldsymbol{\phi}_{\mathbf{b}}$, we get a closed-form expression for the variational beliefs over $\mathbf{B}$, which can be expressed in terms of the Dirichlet prior parameters $\mathbf{b}$ and the variational posterior over hidden states at current and previous timesteps $\boldsymbol{\phi}_{\mathbf{s}_t}$ and $\boldsymbol{\phi}_{\mathbf{s}_{t-1}}$.

\begin{align}
    \mathbf{B}_{t+1} = \frac{\boldsymbol{\phi}_{\mathbf{b}}^*}{\boldsymbol{\phi}_{\mathbf{b}, 0}}
    \label{eq:update_equation}
\end{align}

\begin{align}
    \boldsymbol{\phi}_{\mathbf{b}}^* = \mathbf{b} + \eta (\boldsymbol{\phi}_{\mathbf{s}_t} \otimes \boldsymbol{\phi}_{\mathbf{s}_{t-1}} )
    \label{eq:updateequation}
\end{align}
where \eqref{eq:update_equation} represents the update to the Dirichlet prior for the transition distribution during learning. This is updated with respect to the learning rate $\eta$ and the transition probabilities given the previously performed action $a_{t-1}$. It is this normalised updated Dirichlet prior that then becomes the new transition probability distribution for the following trial.

The updates to the transition model are governed by the sequence of game states. We can imagine a fictive 1-turn sequence (two trials) to imagine how a particular sequence influences learning. If at one trial, the agents both cooperated, then they will infer that the game state was CC. Given this belief, they will infer which action to take. If they choose to defect, hoping that the opponent will cooperate again, the resulting inferred state will be that the optimal action is $\mathbf{u}_t = u^{\mathbf{D}}$, and after the trial they will observe the resulting state, DD. At this point, the agents will update their beliefs about likely transitions (encoded in the $B$ matrix parameters), such that there will be a small incremental increase in the conditional probability of DD, given a past state of CC and a past action of $u^{D}$, i.e., $P(\textbf{s}_{t+1} = \mathrm{DD} |\textbf{s}_t = \mathrm{CC}, \textbf{u}_t = u^{\mathbf{D}})$. The size of this update is determined by a learning rate parameter $\eta$.

\subsection{Deriving the analytic form of the transition function}

When two deterministic agents have the same learning rate, they will perform the same action at every timestep. This has the consequence that the two-agent system will only ever explore two out of four states, namely CC and DD.

The posterior belief can be represented as a vector of its parameters, and in the solution of two identical agents, it can take two possible values, which we denote as $s^{\mathbf{CC}}$ and $s^{\mathbf{DD}}$. Because the likelihood distribution is the identity matrix, these will be maximally precise vectors:

\begin{align}
    s^{\mathbf{CC}} = \begin{bmatrix} 1 \\ 0 \\ 0 \\ 0 \end{bmatrix} \qquad
    s^{\mathbf{DD}} = \begin{bmatrix} 0 \\ 0 \\ 0 \\ 1 \end{bmatrix}
\end{align}

The initial Dirichlet parameters of the prior distribution over the transition model are, for the cooperate and defect-conditioned transitions, respectively,

\begin{align}
    \mathbf{b}_0^{\mathbf{C}} = \begin{bmatrix}
    0.5 & 0.5 & 0.5 & 0.5 \\
    0.5 & 0.5 & 0.5 & 0.5\\
    0 & 0 & 0 & 0 \\
    0 & 0 & 0 & 0
    \end{bmatrix} \qquad
    \mathbf{b}_0^{\mathbf{D}} =\begin{bmatrix}
    0 & 0 & 0 & 0 \\
    0 & 0 & 0 & 0\\
    0.5 & 0.5 & 0.5 & 0.5 \\
    0.5 & 0.5 & 0.5 & 0.5
    \end{bmatrix}
\end{align}

This means that at each timestep, there are four possible updates to the parameters of each agent's variational posterior over the transition model $\phi_{\mathbf{b}}$, given the two variational beliefs a given agent might have ($\mathbf{s}^{\mathbf{CC}}$ and $\mathbf{s}^{\mathbf{DD}}$):

\begin{equation}
    \mathbf{\phi}_{\mathbf{b}_{t+1}}^* = \begin{cases}
      \mathbf{b}_0^{\mathbf{C}} + \eta \cdot (s^{\mathbf{CC}} \otimes s^{\mathbf{CC}})t\\
      \mathbf{b}_0^{\mathbf{D}} + \eta \cdot (s^{\mathbf{DD}} \otimes s^{\mathbf{CC}})t \\
      \mathbf{b}_0^{\mathbf{C}} + \eta \cdot (s^{\mathbf{CC}} \otimes s^{\mathbf{DD}} )t\\
      \mathbf{b}_0^{\mathbf{D}} + \eta \cdot (s^{\mathbf{DD}} \otimes s^{\mathbf{DD}} )t\\
    \end{cases}
\end{equation}

When the agents are both defecting (e.g., in the first timestep when the most likely action is defect), then the update rule for the weights of the Dirichlet parameters of the transition matrix is governed by:

\begin{align}
    \mathbf{\phi}_{\mathbf{b}_{t<\tau_1}}^* = \mathbf{b}^{\mathbf{D}}_0 + \eta \left(s^{\mathbf{DD}} \otimes s^{\mathbf{DD}}\right)t \\
    \mathbf{B}_{t+1 < \tau_1} = \dfrac{\mathbf{\phi}_{\mathbf{b}_{t<\tau_1}}^*}{\mathbf{\phi}_{\mathbf{{b}_{t<\tau_1}},0}^*}
    \label{eq:Btau1}
\end{align}

At some critical time $\tau_1$ the probability of cooperation exceeds that of defection, due to the change in the expected free energies of the two actions $\mathbf{G}_{\tau_1}(u = \mathrm{C}) < \mathbf{G}_{\tau_1}(u = \mathrm{D})$. This triggers the beginning of the so-called ``oscillation period'' (see Section \ref{sec:sim_dynamics} in the main text), where agents periodically oscillate between cooperating and defecting with the same phase. We can expand this condition according to \eqref{eq:efe} into the following form:

\begin{align}
    \mathbf{B}_0^{\mathbf{C}}\cdot s^{\mathbf{DD}}_{\tau_1} \cdot \left( \ln \mathbf{B}_0^{\mathbf{C}} \cdot s^{\mathbf{DD}}_{\tau_1} - \ln \mathbf{C} \right) = \mathbf{B}_{\tau_1}^{\mathbf{D}} \cdot s^{\mathbf{DD}}_{\tau_1} \cdot \left( \ln \mathbf{B^{\mathbf{D}}}_{\tau_1} \cdot s^{\mathbf{DD}}_{\tau_1} - \ln \mathbf{C} \right) \label{eq:oscillationsbegin}
\end{align}

As shown in Section \ref{sec:appendix_tau1}, the equality in \eqref{eq:oscillationsbegin} can be written in terms of $\eta$, $\mathbf{C}$ and $\tau_1$:

\begin{align}
    \frac{1}{(2+\eta 2\tau_1)} \left[ \ln \frac{1}{2(1+\eta \tau_1)\mathbf{C}_3} + (1+ 2 \eta \tau_1) \ln \frac{1+ 2 \eta \tau_1}{2(1+\eta \tau_1)\mathbf{C}_4} \right] = \frac{1}{2} \ln\left(\frac{1}{4\mathbf{C}_1\mathbf{C}_2}\right),
\label{eq:A.32}
\end{align}

Letting $y = \frac{1}{2+2\eta \tau_1}$, which will always be between 0 and 1, we can now rewrite \eqref{eq:A.32} as

\begin{align}
    y \ln y - y \ln \mathbf{C}_3 + (1-y)\ln(1-y) - (1-y) \ln \mathbf{C}_4 &= \frac{1}{2} \ln \left(\frac{1}{4\mathbf{C}_1\mathbf{C}_2} \right)
    \label{eq:ysubstitution}
\end{align}

To derive $\tau_1$ in terms of $\eta$, we must make an approximation. We use the fact that when $y$ is between 0 and 1, it can be approximated by $y \approx Ay^b(y-1) $. This gives us the following expression as an approximation for (35)

\begin{align}
    Ay^b(y-1) - y\ln\mathbf{C}_3 - A(1-y)^by - (1-y)\ln\mathbf{C}_4 = \frac{1}{2} \ln \left(\frac{1}{4\mathbf{C}_1\mathbf{C}_2}\right)
    \label{eq:approximation1}
\end{align}

The optimal values for the approximation are $A = \frac{4774}{4563}$ and $b=\frac{3}{5}$, however, for simplicity, we let $A=1$ and $b=1$ and then the desired root of \eqref{eq:approximation1} can be solved as:

\begin{align}
    y = \frac{1}{4}\big(\ln \frac{\mathbf{C}_3}{\mathbf{C}_4} + 2 - \sqrt{(\ln \frac{\mathbf{C}_4}{\mathbf{C}_3}-2)^2 - 8(-\ln \frac{\mathbf{C}_4}{2\sqrt{\mathbf{C}_1 \mathbf{C}_2}} - \frac{1}{5})} \big)
\end{align}

Therefore, since $y = \frac{1}{2+2\eta \tau_1}$, we have that
\begin{align}
    \tau_1 \approx \frac{R_1}{\eta}
\end{align}
where
\begin{align}
    R_1 = \frac{2}{\ln \frac{\mathbf{C}_3}{\mathbf{C}_4} + 2 - \sqrt{(\ln \frac{\mathbf{C}_4}{\mathbf{C}_3}-2)^2 - 8(-\ln \frac{\mathbf{C}_4}{2\sqrt{\mathbf{C}_1 \mathbf{C}_2}} - \frac{1}{5})}} -1
\end{align}

\begin{figure}[t!]
    \centering
    \includegraphics[width=\textwidth]{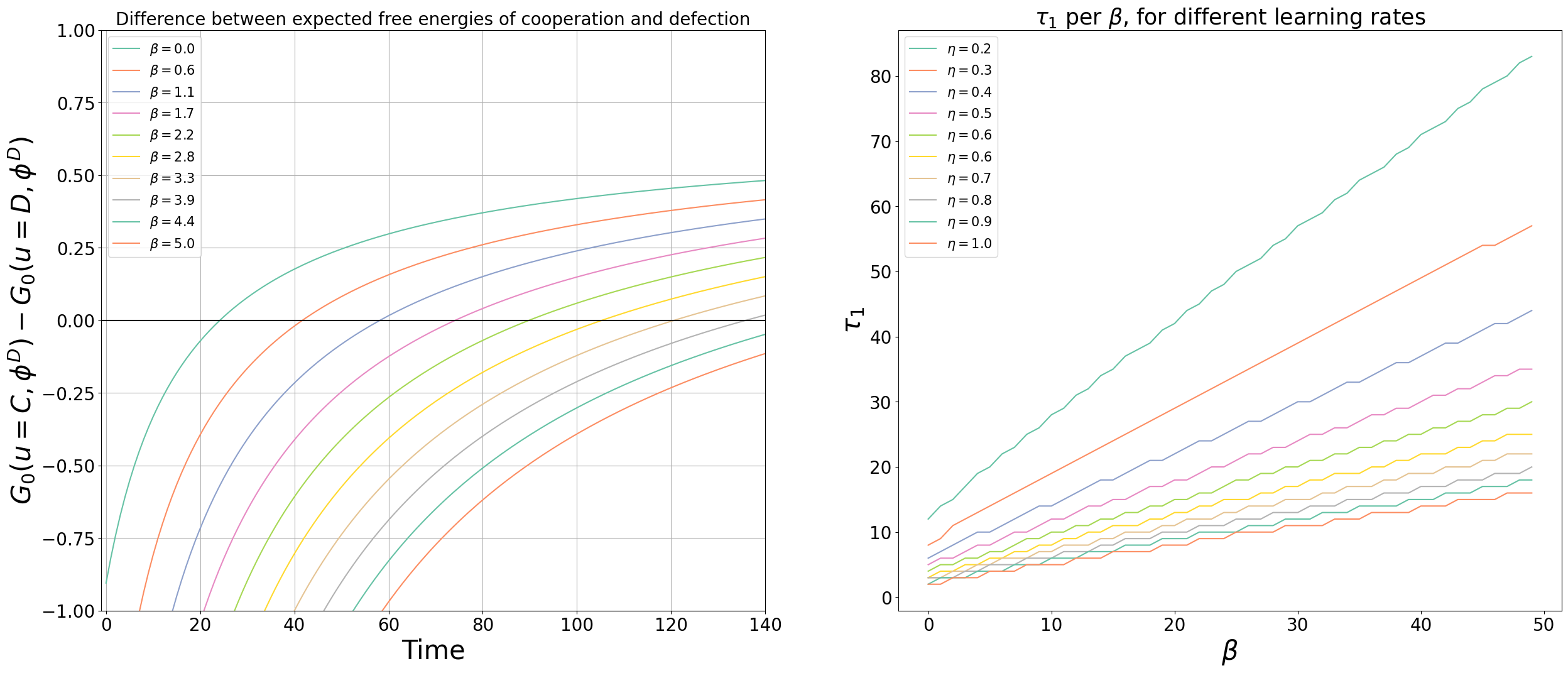}
    \label{fig:difference_of_values_of_efe}
    \caption{\textbf{Dynamics of the expected free energy.} \textbf{Left:} The difference of EFE for cooperation and defection (vertical axis). The roots of this equation are the values of $\tau_1$ for different values of $\beta$, parameterizing the values in the reward function $\mathbf{C}$ as per \eqref{eq:inequality_main}, with $\eta=0.2$. It is clear that with a higher value of $\beta$, it will take agents longer to cooperate, i.e.~$\tau_1$ will be larger, demonstrated by the horizontal translations of the curves as $\beta$ increases. \textbf{Right:} Values of $\tau_1$ for different values of $\beta$ parameterizing the reward function, at different learning rates. Again, we see that as $\beta$ increases, $\tau_1$ increases. We can also see that larger $\eta$ competes with higher $\beta$ to decrease $\tau_1$, as the agents update their transition probability distributions at a higher frequency.}
    \label{fig:diffbeta}
\end{figure}

We now have an approximation for $\tau_1$ in terms of $\eta$ and a constant $R_1$, which depends on the reward \textbf{C} which can be parameterised by $\beta$ according to \eqref{eq:reward}.

\begin{align}
    \tau_1 = \frac{R_1(\beta)}{\eta}
    \label{eq:tau1}
\end{align}
for some precision $\beta$. We can plot this equation for different values of $\beta$ to see how the values in the reward function influence $\tau_1$ (see Figure \ref{fig:diffbeta}).

For $\tau_1 < t < \tau_2$ (i.e., during the period of oscillation dynamics shown in Figure \ref{fig:actionsovertime}), the update rules then become:

\begin{align}
    \mathbf{\phi}_{\mathbf{b}_{\tau_1 < t < \tau_2}}^\mathbf{D} = \mathbf{b}_{\tau_1 }^\mathbf{D} + \frac{1}{2}\eta (s^{\mathbf{DD}} \otimes s^{\mathbf{CC}} )(t - \tau_1)
\label{eq:A.39}
\end{align}
\begin{align}
    \mathbf{\phi}_{\mathbf{b}_{\tau_1 < t < \tau_2}}^\mathbf{C} = \mathbf{b}_0^{\mathbf{C}} + \frac{1}{2}\eta (s^{\mathbf{CC}} \otimes s^{\mathbf{DD}} )(t - \tau_1)
\label{eq:A.40}
\end{align}

The update rule changes from \eqref{eq:A.39} to \eqref{eq:A.40} at every other trial, from conditioning on the previous action being D, to being C. The oscillation period persists until some time $\tau_2$. At $\tau_2$ we will have that, for the first time, $\mathbf{G}_0(u = \mathrm{C}, \phi^{\mathrm{C}}) < \mathbf{G}_0(u = \mathrm{C}, \phi^{\mathrm{D}})$. Again, we can expand this according to \eqref{eq:efe} as:

\begin{align}
    \mathbf{B}_{\tau_2}^{\mathbf{C}}\cdot s^{\mathbf{CC}} \cdot \left( \ln \mathbf{B}_{\tau_2}^{\mathbf{C}} \cdot s^{\mathbf{CC}} - \ln \mathbf{C} \right) = \mathbf{B}_{\tau_2}^{\mathbf{D}} \cdot s^{\mathbf{CC}} \cdot \left( \ln \mathbf{B^{\mathbf{D}}}_{\tau_1} \cdot s^{\mathbf{CC}} - \ln \mathbf{C} \right)
\end{align}

Rewriting this equation in terms of $\eta$, $\tau_1$, $\tau_2$, and $\mathbf{C}$ leads to the following inequality (for full derivation, see Section \ref{sec:appendix_tau2}):

\begin{align}
    \frac{1}{2 + \eta(\tau_2-\tau_1)} \left[\ln [\frac{1}{\mathbf{C}_3}(\frac{1}{2+\eta(\tau_2-\tau_1)})] + (1 + \eta(\tau_2-\tau_1))\ln\frac{1 + \eta(\tau_2-\tau_1)}{\mathbf{C}_4(2 + \eta(\tau_2-\tau_1))} \right] = - \frac{1}{2}\ln(4\mathbf{C}_1\mathbf{C}_2) \label{eq:inequality_main}
\end{align}

This time, we let $y=\frac{1}{2+\eta(\tau_2-\tau_1)}$ and we have:

\begin{align}
    (y-1) \ln y - y\ln \mathbf{C}_3 + (1-y)\ln(1-y) - (1-y)\ln\mathbf{C}_4 = - \frac{1}{2}\ln(4\mathbf{C}_1\mathbf{C}_2)
\end{align}

Now, notice that this is the exact same equation as \eqref{eq:ysubstitution} above, which we know we can approximate as \eqref{eq:approximation1}. We can then write our solution in terms of $R_1$:

\begin{align}
     \tau_2 \approx \frac{1}{\eta}(\frac{1}{y} -2) + \tau_1 = \frac{1}{\eta} (\frac{3}{2}R_1)
\end{align}

The resulting equation is obtained in terms of $R_2$, where $R_2 = \frac{3}{2}R_1$.

\begin{align}
    \tau_2 \approx \frac{R_2(\beta)}{\eta}
\end{align}

After $\tau_2$, agents will cooperate indefinitely according to the final steady state update rule:

\begin{align}
    \mathbf{\phi}_{\mathbf{b}_{t>\tau_2}}^* = \mathbf{b}^{\mathbf{C}}_{\tau_2} + \eta \left(s^{\mathbf{CC}} \otimes s^{\mathbf{CC}}\right)t \\
    \mathbf{B}_{t+1 > \tau_2} = \frac{\mathbf{\phi}_{\mathbf{b}_{t>\tau_2}}^*}{\mathbf{\phi}_{\mathbf{{b}_{t>\tau_2}},0}^*}
    \label{eq:Btau2}
\end{align}

\subsection{Full derivation of \texorpdfstring{$\tau_1$}{Lg}}\label{sec:appendix_tau1}

Here we derive $\tau_1$ for the following equality from \eqref{eq:oscillationsbegin}:

\begin{align}
    \mathbf{B}_0^{\mathbf{C}}\cdot s^{\mathbf{DD}}_{\tau_1} \cdot \left( \ln \mathbf{B}_0^{\mathbf{C}} \cdot s^{\mathbf{DD}}_{\tau_1} - \ln \mathbf{C} \right) = \mathbf{B}_{\tau_1}^{\mathbf{D}} \cdot s^{\mathbf{DD}}_{\tau_1} \cdot \left( \ln \mathbf{B^{\mathbf{D}}}_{\tau_1} \cdot s^{\mathbf{DD}}_{\tau_1} - \ln \mathbf{C} \right)
\end{align}

Using the following:

\begin{align}
    \mathbf{B}_{t} = \frac{\mathbf{\phi}_{\mathbf{b}_{t}}}{\mathbf{\phi}_{\mathbf{{b}_{t},0}}} \\
    \mathbf{\phi}_{\mathbf{b}_{t<\tau_1}}^{\mathbf{D}} = \mathbf{b}^{\mathbf{D}}_0 + \eta (s^{\mathbf{DD}} \otimes s^{\mathbf{DD}} )t \\
    \mathbf{\phi}_{\mathbf{b}_{t<\tau_1}}^{\mathbf{C}} = \mathbf{b}_0^{\mathbf{C}} \\
    \mathbf{b}_0^{\mathbf{C}} = \begin{bmatrix}
    0.5 & 0.5 & 0.5 & 0.5 \\
    0.5 & 0.5 & 0.5 & 0.5\\
    0 & 0 & 0 & 0 \\
    0 & 0 & 0 & 0
    \end{bmatrix} \qquad
    \mathbf{b}_0^{\mathbf{D}} =\begin{bmatrix}
    0 & 0 & 0 & 0 \\
    0 & 0 & 0 & 0\\
    0.5 & 0.5 & 0.5 & 0.5 \\
    0.5 & 0.5 & 0.5 & 0.5
    \end{bmatrix} \\
    s^{\mathbf{DD}} = \mathbf{e}_4,
\end{align}
we have:

\begin{align}
    \frac{\mathbf{\phi}_{\mathbf{b}_{0}}^\mathbf{C}}{\mathbf{\phi}_{\mathbf{{b}_{0}},0}^\mathbf{C}} \cdot s^\mathbf{DD} \cdot \left( \ln \frac{\mathbf{\phi}_{\mathbf{b}_{0}}^\mathbf{C}}{\mathbf{\phi}_{\mathbf{{b}_{0}},0}^\mathbf{C}} \cdot s^\mathbf{DD} - \ln \mathbf{C} \right) = \frac{\mathbf{\phi}_{\mathbf{b}_{\tau_1}}^\mathbf{D}}{\mathbf{\phi}_{\mathbf{{b}_{\tau_1}},0}^\mathbf{D}} \cdot s^\mathbf{DD} \cdot \left( \ln \frac{\mathbf{\phi}_{\mathbf{b}_{\tau_1}}^\mathbf{D}}{\mathbf{\phi}_{\mathbf{{b}_{\tau_1}},0}^\mathbf{D}} \cdot s^\mathbf{DD} - \ln \mathbf{C} \right)
\end{align}

On the LHS:

\begin{align}
    \frac{\mathbf{\phi}_{\mathbf{b}_{0}}^\mathbf{C}}{\mathbf{\phi}_{\mathbf{{b}_{0}},0}^\mathbf{C}} \cdot s^\mathbf{DD} \cdot \left( \ln \frac{\mathbf{\phi}_{\mathbf{b}_{0}}^\mathbf{C}}{\mathbf{\phi}_{\mathbf{{b}_{0}},0}^\mathbf{C}} \cdot s^\mathbf{DD} - \ln \mathbf{C} \right) = (\mathbf{b}_0^{\mathbf{C}} \cdot \mathbf{e}_4) \cdot \ln \frac{\mathbf{b}_0^{\mathbf{C}} \cdot \mathbf{e}_4}{\mathbf{C}} = -\frac{1}{2}\ln(4\mathbf{C}_1\mathbf{C}_2)
\end{align}

On the RHS:

\begin{align}
    \frac{\mathbf{\phi}_{\mathbf{b}_{\tau_1}}^\mathbf{D}}{\mathbf{\phi}_{\mathbf{{b}_{\tau_1}},0}^\mathbf{D}} \cdot s^\mathbf{DD} \cdot \left( \ln \frac{\mathbf{\phi}_{\mathbf{b}_{\tau_1}}^\mathbf{D}}{\mathbf{\phi}_{\mathbf{{b}_{\tau_1}},0}^\mathbf{D}} \cdot s^\mathbf{DD} - \ln \mathbf{C} \right) = \frac{\mathbf{\phi}_{\mathbf{b}_{\tau_1}^\mathbf{D}, j=4}}{\mathbf{\phi}_{\mathbf{{b}_{\tau_1}},0}^\mathbf{D}} \cdot \left(\ln \frac{\mathbf{\phi}_{\mathbf{b}_{\tau_1}^\mathbf{D}, j=4}}{\mathbf{\phi}_{\mathbf{{b}_{\tau_1}},0}^\mathbf{D}} - \ln \mathbf{C} \right) \\
    = \frac{1}{2}\frac{1}{1+\eta \tau_1} \ln (\frac{1}{2(1+\eta \tau_1)\mathbf{C}_3}) + \frac{1}{2} \frac{1+2\eta\tau_1}{1+\eta\tau_1}\ln (\frac{1 + \eta \tau_1}{2(1+\eta\tau_1)\mathbf{C}_4})
\end{align}

Our equality is therefore:

\begin{align}
    \frac{1}{(2+2\eta \tau_1)} \left[ \ln \frac{1}{2(1+\eta \tau_1)\mathbf{C}_3} + (1+ 2 \eta \tau_1) \ln \frac{1+ 2 \eta \tau_1}{2(1+\eta \tau_1)\mathbf{C}_4} \right] = - \frac{1}{2}\ln(4\mathbf{C}_1\mathbf{C}_2)
\end{align}

\subsection{Full derivation of \texorpdfstring{$\tau_2$}{Lg}}\label{sec:appendix_tau2}

Our condition for deriving $\tau_2$ in terms of the expected free energies is

\begin{align}
    \mathbf{B}_{\tau_2}^{\mathbf{C}}\cdot s^{\mathbf{CC}} \cdot \left( \ln \mathbf{B}_{\tau_2}^{\mathbf{C}} \cdot s^{\mathbf{CC}} - \ln \mathbf{C} \right) = \mathbf{B}_{\tau_2}^{\mathbf{D}} \cdot s^{\mathbf{CC}} \cdot \left( \ln \mathbf{B^{\mathbf{D}}}_{\tau_2} \cdot s^{\mathbf{CC}} - \ln \mathbf{C} \right)
\end{align}

Here our $\phi$s between trials $\tau_1$ and $\tau_2$ are:

\begin{align}
    \mathbf{\phi}_{\mathbf{b}_{\tau_1 < t < \tau_2}}^\mathbf{D} = \mathbf{\phi}_{\mathbf{b}_{t < \tau_1 }}^\mathbf{D} + \frac{1}{2}\eta (s^{\mathbf{DD}} \otimes s^{\mathbf{CC}} )(t - \tau_1) \\
    \mathbf{\phi}_{\mathbf{b}_{\tau_1 < t < \tau_2}}^\mathbf{C} = \mathbf{b}_0^{\mathbf{C}} + \frac{1}{2}\eta (s^{\mathbf{CC}} \otimes s^{\mathbf{DD}} )(t - \tau_1)
\end{align}

And to solve for $\tau_2$ our inequality is

\begin{align}
    \frac{\mathbf{\phi}_{\mathbf{b}_{\tau_2}}^\mathbf{C}}{\mathbf{\phi}_{\mathbf{b}_{\tau_2}, 0}^\mathbf{C}} \cdot s^\mathbf{CC} \cdot \left( \ln \frac{\mathbf{\phi}_{\mathbf{b}_{\tau_2}}^\mathbf{C}}{\mathbf{\phi}_{\mathbf{b}_{\tau_2}, 0}^\mathbf{C}} \cdot s^\mathbf{CC} - \ln \mathbf{C} \right) = \frac{\mathbf{\phi}_{\mathbf{b}_{\tau_2}}^\mathbf{D}}{\mathbf{\phi}_{\mathbf{b}_{\tau_2},0}^\mathbf{D}} \cdot s^\mathbf{DD} \cdot \left( \ln \frac{\mathbf{\phi}_{\mathbf{b}_{\tau_2}}^\mathbf{D}}{\mathbf{\phi}_{\mathbf{b}_{\tau_2},0}^\mathbf{D}} \cdot s^\mathbf{DD} - \ln \mathbf{C} \right)
\end{align}

On the LHS we have:

\begin{align}
    \frac{\mathbf{\phi}_{\mathbf{b}_{\tau_2}}^\mathbf{C}}{\mathbf{\phi}_{\mathbf{b}_{\tau_2}, 0}^\mathbf{C}} \cdot s^\mathbf{CC} \cdot \left( \ln \frac{\mathbf{\phi}_{\mathbf{b}_{\tau_2}}^\mathbf{C}}{\mathbf{\phi}_{\mathbf{b}_{\tau_2}, 0}^\mathbf{C}} \cdot s^\mathbf{CC} - \ln \mathbf{C} \right) = - \frac{1}{2}\ln(4\mathbf{C}_1\mathbf{C}_2)
\end{align}

On the RHS:

\begin{align}
    \frac{\mathbf{\phi}_{\mathbf{b}_{\tau_2}}^\mathbf{C}}{\mathbf{\phi}_{\mathbf{b}_{\tau_2},0}^\mathbf{C}} \cdot s^\mathbf{DD} = \frac{1}{2 + \eta(\tau_2-\tau_1)}\begin{bmatrix} 0 \\ 0 \\ 1 \\1+\eta(\tau_2 - \tau_1) \end{bmatrix} \\\notag\\
    \frac{1}{2 + \eta(\tau_2-\tau_1)} \left[\ln \left[\frac{1}{\mathbf{C}_3}(\frac{1}{2+\eta(\tau_2-\tau_1)})\right] + (1 + \eta(\tau_2-\tau_1))\ln\frac{1 + \eta(\tau_2-\tau_1)}{\mathbf{C}_4(2 + \eta(\tau_2-\tau_1))} \right] \\\notag\\
    \frac{1}{2 + \eta(\tau_2-\tau_1)} \ln \left[\frac{\mathbf{C}_4}{\mathbf{C}_3(1+\eta(\tau_2-\tau_1)})\right] + \ln\frac{1 + \eta(\tau_2-\tau_1)}{\mathbf{C}_4(2 + \eta(\tau_2-\tau_1))}
\end{align}

Finally, our inequality is:

\begin{align}
    \frac{1}{2 + \eta(\tau_2-\tau_1)} \Bigg[ & \ln \left[\frac{1}{\mathbf{C}_3}\left(\frac{1}{2+\eta(\tau_2-\tau_1)}\right)\right] + \notag\\
    & (1 + \eta(\tau_2-\tau_1))\ln\frac{1 + \eta(\tau_2-\tau_1)}{\mathbf{C}_4(2 + \eta(\tau_2-\tau_1))} \Bigg] = - \frac{1}{2}\ln(4\mathbf{C}_1\mathbf{C}_2)
\end{align}

\begin{sloppypar}
\printbibliography[title={Supplemental References}]
\end{sloppypar}
\end{refsection}

\end{document}